\documentclass[aps,prb,showpacs,12pt]{revtex4}
\usepackage{graphicx}

\newcommand{\beq}{\begin{equation}}
\newcommand{\eeq}{\end{equation}}

\begin{document}

\title{Stability of a Bose-Einstein condensate revisited for composite bosons}

\author{M. Combescot}

\affiliation{Institut des NanoSciences de Paris, CNRS, Universite Pierre et Marie Curie\\
140 rue de Lourmel, 75015 Paris, France}

\author{D.W. Snoke}
\affiliation{Department of Physics and Astronomy, University of Pittsburgh\\
3941 O'Hara St., Pittsburgh, PA 15260, USA}

%\date{}

\begin{abstract}
It is known that elementary bosons condense in a unique state, not so much because this state has the lowest free particle energy but because it costs a macroscopic amount of energy to put the particles into different states which can then interact through quantum particle exchanges. Since individual exchanges between the two fermions of a composite boson are ignored when composite particles are replaced by elementary bosons, it is of importance to reconsider the exchange energy argument for the stability of the Bose-Einstein condensate in the case of composite bosons. We do this here in the light of the new many-body theory which allows us to take exactly into account all possible exchanges between the fermionic components of the composite bosons. We confirm that the condensate of composite bosons is made of a unique state, this state being moreover {\em pure}: a coherent superposition of states close in energy is shown to be less favorable for both elementary and composite bosons.
\end{abstract}

\maketitle

\section{Introduction}

Bose-Einstein Condensation (BEC) is of high current interest due to recent observations of condensates made of atoms \cite{r1}, molecules \cite{r2,r3,r3a,r3b}, and polaritons \cite{r4,r5}, which result from the strong coupling of a photon and an exciton \cite{r6}. The experimental observation of a pure exciton condensate \cite{r6a,r6b}, however, remains a challenge. It has been recently argued \cite{r7} that the condensate in this system should appear in a dark state, since these are the lowest energy states due to the weak valence-conduction repulsive processes which do not exist for excitons with spin S=(+2,-2). Consequently, such a dark condensate cannot be directly seen by optical emission investigated up to now \cite{r8}, though it should be possible to deduce its presence indirectly. In addition, excitons are excited states; so that, in order to reach condensation, excitons with lifetime long compared to their thermalization time are needed. This has motivated the development of coupled quantum well structures \cite{r11} with electrons well separated from holes to increase their recombination time.

The standard formulation of BEC considers a set of free elementary bosons. For particles with a center of mass momentum, a special role is then played by the zero momentum ($\vec{k}=0$) state. However, since the energy spectrum of such particles confined in a large volume is essentially continuous, it is physically hard  to accept that this lowest energy state is favored over all the other nearby states just on account of an infinitesimally small kinetic energy difference. 

Actually, the essential characteristic of a Bose-Einstein condensate, that there is a macroscopic number of particles in a single quantum state, arises not from an argument for an ideal gas but from interactions. This has been made in a very straightforward way by Nozi\`eres \cite{r12,r13}: By calculating the energy in the Born approximation of $N$ elementary bosons with repulsive interaction (necessary to avoid a density collapse) he has shown that to break up the condensate into two different states, we must pay a macroscopic exchange energy penalty which increases as $N^2$.   As Nozi\`eres said, "...it is the exchange interaction energy that makes condensation fragmentation costly.  Genuine Bose-Einstein condensation is not an ideal gas effect: it implies interacting particles!"

This nicely shows that exchange between undistinguishable quantum particles plays a fundamental, not peripheral, role in the essential physics of Bose-Einstein condensation.  A similar role is actually played in the spin-spin $J$-coupling  of ferromagnets, favoring particles to be in the same state: Without this coupling, there is no ferromagnetic phase transition.   As shown below, even for just $N=2$ bosons, the result is still valid; in other words, there already is an exchange energy penalty for two bosons to be in different states.  As a general rule, we can say that Bose-Einstein condensation occurs when the exchange energy for two bosons becomes comparable to their thermal energy. 

As Nozi\`eres' conclusion was obtained before the development of the composite boson many-body theory, it was  de facto reached in the framework of elementary bosons. Yet, all real condensates consist of composite bosons which are made of an even number of fermions. Due to the Pauli exclusion principle between fermions, additional exchange processes between the fermions which make up the composite bosons must be considered in the overall exchange energy.  These effects are expected to be important not only for excitons composed of light-mass electron and hole, but also for atoms, since the electron exchange energy is known to enter the effective interaction between two atoms. 

A number of works\cite{t1,t2,t3,t4,t5,t6,t7,t8,t9,t10,t11} have addressed interaction and condensation of composite bosons (``cobosons" in short) such as excitons. One approach is to stay completely in the fermion picture. Treating the correlated pairs in this approach can be cumbersome and can require heavy numerical methods.   Another popular approach is the method of bosonization, in which the fermion pairs are treated as pure bosons but with an altered interaction taking into account the underlying Fermi statistics. As discussed in a series of recent papers presenting a new approach to composite boson theory\cite{r14,r15}, the bosonization method neglects certain exchange processes which are important even in low-order perturbation theory, and neglecting these can sometimes have dramatic consequences, for example, neglecting dominant terms in semiconductor optical nonlinearities \cite{r22,r22a}.  In the present paper we will use the new composite boson theory, which has a convenient diagrammatic method which lends itself to analytical results. We reconsider the overall stability of Bose-Einstein condensation in the case of composite bosons using the framework of this new many-body theory, since exchange, which is crucial for the condensate stability, is not properly treated when the fermionic components of the particles are forgotten.  Our aim in this paper is to show that, indeed, the exchange-energy stability argument still applies in the case of composite bosons, using the new tools that this composite boson many-body theory now offers. 
 
This many-body theory shows that two composite bosons interact through two conceptually different scatterings \cite{r14}:  ``interaction scatterings" for fermion interactions in the absence of fermion exchange, which have energylike quantities, and dimensionless ``Pauli scatterings''  for fermion exchanges in the absence of fermion interaction. These Pauli scatterings, by construction ignored when the composite particles are ``bosonized", turn out to be crucial in the many-body physics of composite bosons: they, in particular, control all semiconductor optical nonlinearities induced by unabsorbed photons. From these $2\times 2$ Pauli scatterings, we can construct any possible fermion exchange which exists between $N$ composite bosons, as necessary since the Pauli exclusion principle from which these exchanges originate, is $N$-body by essence. These $N$-body exchanges are nicely visualized through new diagrams, called ``Shiva diagrams" \cite{r17}, which not only allow one to see the subtle many-body physics taking place between these tricky objects but also to calculate it readily.
 
 This new formalism also shows that it is impossible to write an effective Hamiltonian for bosonized excitons which produces the correct scattering rates and the correct lifetime of $N$ exciton states \cite{r18}, even in the extreme dilute limit of $N=2$.  A way to grasp the difficulty is to note that, by mapping composite bosons into an elementary boson subspace, we strongly reduce the degrees of freedom of the problem.  This mathematically shows up through a change from $1/N!$ to $(1/N!)^2$ in the prefactor of the closure relation for $N$ elementary or composite bosons \cite{r19}, making all sum rules irretrievably different.

In this paper, we consider the possibility to have not only a condensate made of two different states but also a condensate made of a coherent superposition of states close in energy, for both elementary and composite bosons.  Thus, we are going to consider the three states,
\begin{eqnarray}
&& |{\phi}_0\rangle =  {B}_o^{{\dagger N}}|0\rangle, \label{1-1}\\
 &&|{\phi}_{12}\rangle = {B}_{o_1}^{\dagger {N_1}} {B}_{o_2}^{\dagger {N_2}}|0\rangle,\\
&&|{\phi}\rangle = (a'{B}^\dagger_{o'}+ a''{B}^\dagger _{o''})^N|0\rangle,\label{newstate}
\end{eqnarray} 
where the ${B}^\dagger _i$ operators are creation operators for composite bosons, and we will compare the results with the ones obtained with elementary boson operators, written with a bar as $\bar{B}^{\dagger}_i$.  These operators having the standard commutation relation $[\bar{B}_m,\bar{B}^\dagger _i] = \delta_{mi}$.  For composite bosons, the ${B}^\dagger _i$ operators obey  the relation $[{B}_m,{B}^\dagger _i] = \delta_{mi} - D_{mi}$, where the ``deviation-from-boson'' operator  $D_{mi}$ will be discussed in more detail below. The third state, $|\phi\rangle$, defined in (\ref{newstate}) is a coherent superposition of composite bosons in two different eigenstates. The possibility to have such a coherent superposition, not considered by Nozi\`eres but considered by Leggett \cite{r13}, will also be addressed here in the case of elementary bosons. In the spirit of the Nozi\`eres and Leggett arguments, we consider only the lowest order of perturbation theory, calculating the mean-field energy of the different states.  This is certainly valid in the low-density limit, although as discussed by Leggett \cite{r13}, a rigorous demonstration of the stability of a condensate in the general case of interacting bosons at all densities does not yet exist. We confirm that the mean value of the Hamiltonian is minimum for $|\bar{\phi}_0\rangle$ and $|\phi_0\rangle$; i.e., for a condensate which is pure and not fragmented, whatever the repulsive scatterings between particles, and whether or not the bosons are elementary or composite.
 
As the calculations for a large number of composite bosons are heavy and quite technical, especially for readers who are not yet familiar with this new many-body theory for composite bosons, we first perform the calculations for $N=2$.  Most of the important physics is usually seen just from examining two interacting particles.  A nice rule of thumb---which will be explicitly confirmed---allows us to obtain the results for $N$ by replacing $(a_B/L)^d$ in the results for $N=2$ by
\begin{equation}
\eta = N(a_B/L)^d,
\label{eta}
\end{equation}
where $a_B$ is the composite boson wavefunction extension, $L$ the sample size, and $d$ the spatial dimension.  This rule of thumb is less obvious for the fragmented state $|\phi_{12}\rangle$ because we then have two boson populations $N_1$ and $N_2$. However, we physically expect to have the exchange terms between the composite bosons in states 1 and 2 to appear with a prefactor $N_1N_2$, while the ones involving exchange only in state 1 or only in state 2 should appear with prefactors $N_1(N_1-1)/2$ and $N_2(N_2-1)/2$ for the number of ways to choose two particles among the $N_1$ or $N_2$ populations.
 
As with other many-body effects between composite bosons, the tricky part of the calculation always is to determine the density expansion of the relevant scalar products for $N$ composite boson states.  While it is always possible to get them through a brute force algebra making use of the commutations \cite{r14} on which the composite boson many body theory is based, these expansions are nicely performed by using Shiva diagrams \cite{r17} which visualize fermion exchanges in a transparent way: the density expansions are naturally associated with diagrams having an increasing number of composite boson lines.  In the case of state $|\phi_{12}\rangle$, however, we must work a little harder, because we now have two large numbers which are relevant, $N_1$ and $N_2$.  In order to determine the extra energy due to the fracturing of the condensate into two different states, we must carefully distinguish between the exchanges which are internal to the $N_1$ population, those which are internal to the $N_2$ population, and those which occur between these two populations.  Only these last contribute to the exchange penalty for fragmenting the condensate into two different states. 
 
The paper is organized as follows:

In Sections \ref{sect.form1} and \ref{sect.form2}, we review the formalism for many-body effects with elementary and composite bosons, and settle the notations. An important part of this discussion is the normalization of $N$-particle states, which is trivial for elementary bosons but far from trivial for composite bosons. In particular, we present new results for the normalization of a state made of two large numbers of different cobosons. 

In Section \ref{sect.two1}, we calculate the mean value of an effective Hamiltonian describing interacting elementary bosons, in the case that of {\em two} elementary bosons in each of the three configurations $|\bar{\phi}_0\rangle,~|\bar{\phi}_{12}\rangle$ and $|\bar{\phi}\rangle$, while in Section \ref{sect.two2} we perform the same calculations for two {\em composite} bosons, using the exact Hamiltonian for interacting fermions. This in particular, allows us to identify the proper effective scatterings one has to take for diagonal processes between bosonized particles in terms of the interaction and Pauli scatterings appearing in the composite boson many-body theory.

In Section \ref{sect.many1}, we calculate the effective Hamiltonian mean value for $N$ elementary bosons in the three states of interest, namely pure, fragmented and coherent, while in Section \ref{sect.many2}, we address the same problem for $N$ composite bosons.

In Section \ref{sect.concl}, we discuss the results and conclude.

In the appendix, we briefly reproduce  Nozi\`eres's original argument, for completeness, and also because the spirit of the present paper is the same: by calculating the Hamiltonian mean values in the states $|\bar{\phi}_0\rangle,~|\bar{\phi}_{12}\rangle$ and $|\bar{\phi}\rangle$, we by construction study the effect of interactions between bosons in the Born approximation, i.e., to first order in the interactions. States $\bar{B}^{\dagger N}_0|0\rangle$ and ${B}^{\dagger N}_0|0\rangle$ are the system ground states in zero order for the elementary and composite bosons, respectively. For those states, the mean value of the Hamiltonian reduces to $NE_0$ in the low-density limit. 
By drawing our conclusion  about the non-fragmentation of the condensate from the sign of the term linear in density, obtained within the Born approximation, we of course implicitly assume that higher-order terms in density will not modify the overall sign, which is likely in view of our past knowledge of many-body effects. This approach assumes zero temperature and low density.  Actually, low density is implicitly assumed when considering excitons, which can convert to electron-hole plasma as density increases. \cite{snoke-ion}

\section{Formalism for elementary bosons}
\label{sect.form1}
\setcounter{equation}{0}

\begin{figure}
\begin{center}
\includegraphics[width=0.29\textwidth]{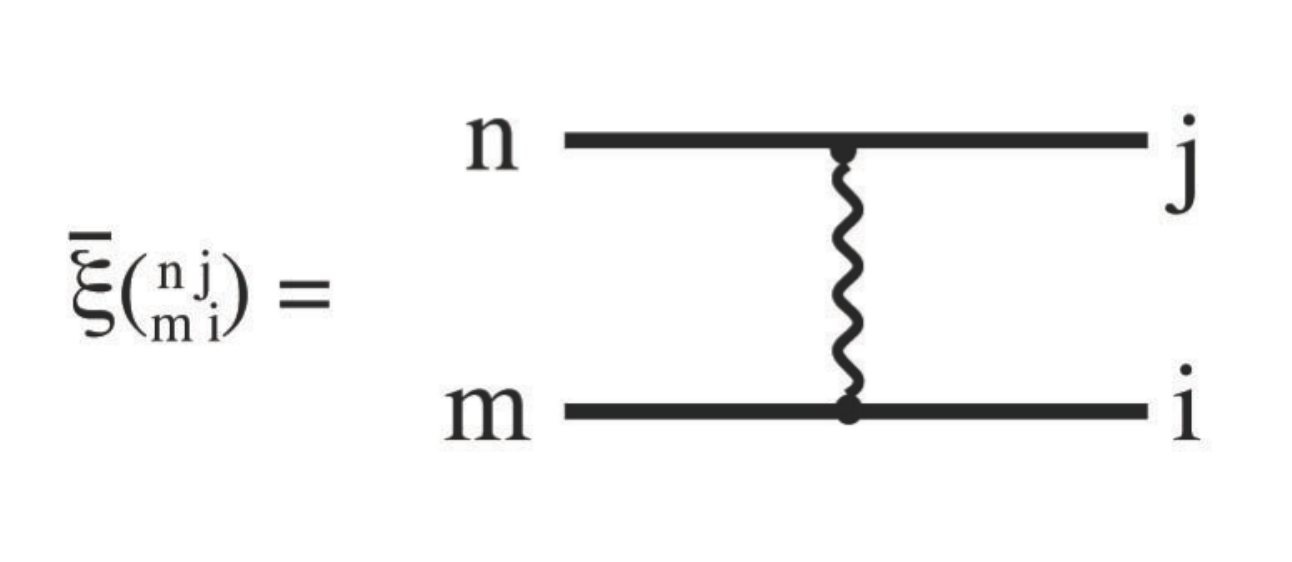}
\end{center}
\caption{Scattering $\bar\xi(^n_m \left.\right.^j_i)$, as defined in equation (\ref{eV}), between two elementary bosons.}
\label{fig1}
\end{figure}

In this section, we briefly recall the formalism for elementary-boson many-body calculations. The commutation relations for these bosons are
 \begin{equation}
 [\bar{B}_m,\bar{B}_i] = 0, \hspace{1cm}  [\bar{B}_m,\bar{B}^\dagger_i ] = \delta_{mi}.
\end{equation}
The index $i$ usually stands for a center-of-mass momentum $\vec{Q}_i$ and a relative motion index ${\nu}_i$. For elementary bosons, it is possible to split the system Hamiltonian as $\bar{H} = \bar{H}_0+\bar{V}$ where the one-body part is  $\bar{H}_0 = \sum{E_i\bar{B}^\dagger_i\bar{B}_i}$ while the interaction can be written as
\begin{equation}
\bar{V} = \frac{1}{2} \sum{\bar{\xi}
(^n_m \left. \right.^j_i  )\bar{B}^\dagger_m \bar{B}^\dagger_n \bar{B}_j\bar{B}_i}.
\label{eV}
\end{equation}
The energy-like prefactor 
$\displaystyle \bar{\xi} (^n_m \left. \right.^j_i  )$, which describes the scattering from $i$ to $m$ and $j$ to $n$, (see Fig.~\ref{fig1}), must be such that $\displaystyle\bar{\xi} (^n_m \left. \right.^j_i )^*  = \bar{\xi}(^j_i\left. \right.^n_m  )$ to insure hermiticity. Momentum conservation requires that this scattering differs from zero only for $\vec{Q}_m+\vec{Q}_n = \vec{Q}_i+\vec{Q}_j$.  Nozi\`eres' argument \cite{r12} was made for a structureless scattering, i.e., for $\bar{\xi}(^n_m \left. \right.^j_i  ) = V_0 \delta_{\vec{Q}_m+\vec{Q}_n,\vec{Q}_i+\vec{Q}_j}$. By noting that $\bar{B}^\dagger_m\bar{B}^\dagger_n = \bar{B}^\dagger_n\bar{B}^\dagger_m$, it is possible to rewrite this potential $\bar{V}$ in a fully symmetrical form
\begin{equation} 
\bar{V} = \frac{1}{2}\sum{\bar{\xi}_{mn;ij}\bar{B}^\dagger _m\bar{B}^\dagger _n\bar{B}_j\bar{B}_i},
\end{equation}
where  $\displaystyle \bar{\xi}_{mn;ij}$, equal to $\left[ \bar{\xi} (^n_m \left. \right.^j_i  )+ \bar{\xi} (^m_n \left. \right.^j_i  )\right]/2$, is now such that $\bar{\xi}_{mn;ij} = \bar{\xi}_{mn;ji} = \bar{\xi}_{nm;ij} = \bar{\xi}^*_{ij;mn}$. 

In order to perform many-body calculations with elementary bosons in a convenient way, let us note that
\begin{equation} 
[\bar{B}_m,\bar{B}^{\dagger N}_o] = [\bar{B}_m,\bar{B}^\dagger _o]\bar{B}^{\dagger  N-1}_o + \bar{B}^\dagger _o[\bar{B}_m,\bar{B}^\dagger _o]\bar{B}^{\dagger  N-2}_o + ... = N\delta_{mo}\bar{B}^{\dagger  N-1}_o,
\label{mebB}
\end{equation}
This gives for the state $|\bar{\phi_o}\rangle$ defined in (\ref{1-1})
\begin{equation}
\bar{B}_j\bar{B}_i|\bar{\phi_0}\rangle  = N(N - 1)\delta_{io}\delta_{jo}\bar{B}^{\dagger  N-2}_o|0\rangle .
\end{equation}
 If instead of a single state, we now consider a coherent superposition of two states, $\bar{B}^{\dagger} = a'\bar{B}^{\dagger}_{o'} + a''\bar{B}^{\dagger}_{o''}$, equation (\ref{mebB}) implies 
\begin{equation}
[\bar{B}_i,\bar{B}^{\dagger N}] = N\delta_i\bar{B}^{\dagger N-1}.
\label{cohelmany}
\end{equation}
with $\delta_i$ such that $[\bar{B}_i,\bar{B}^{\dagger}] =\delta_i$, i.e.,
\begin{equation}
\delta_i = a'\delta_{io'} + a''\delta_{io''} 
\label{2-7}
\end{equation}

\section{Formalism for composite bosons}
\label{sect.form2}
\setcounter{equation}{0}

\subsection{Elementary scattering}
 
We now consider composite bosons made of one fermion $\alpha$ and one fermion $\beta$, and we briefly review the formalism presented in Refs. \onlinecite{r14} and \onlinecite{r15}.  The Hamiltonian of these fermions reads
\begin{equation}
H = H_{\alpha} + H_{\beta} + V_{\alpha\alpha} + V_{\beta\beta} + V_{\alpha\beta}.
\end{equation}
 For excitons or hydrogen atoms, $V_{\alpha\alpha}, V_{\beta\beta}$ and $V_{\alpha\beta}$ are Coulomb potentials, while for the so-called cold Fermi gases, $V_{\alpha\alpha} \approx V_{\beta\beta}\approx0$ while $V_{\alpha\beta}$ is short range. These interaction terms can be formally written as
 \begin{equation}
 V_{\alpha \beta} = \sum {\cal V}\left(^{\vec{k}'_\beta}_{\vec{k}'_\alpha} \left. \right.^{\vec{k}^{ }_\beta}_{\vec{k}^{ }_\alpha}  \right)a^{\dagger}_{\vec{k}'_\alpha} b^{\dagger}_{\vec{k}'_\beta}b^{ }_{\vec{k}^{ }_\beta}a^{ }_{\vec{k}^{ }_\alpha}
 \end{equation}
 and similarly for  $V_{\alpha \alpha}$ and $V_{\beta \beta}$ with a $1/2$ prefactor. $a^\dagger _{\vec{k}_{\alpha}}$ and $b^\dagger _{\vec{k}_{\beta}}$ are, for simplicity, chosen to be the creation operators for one fermion $\alpha$ or $\beta$ in an eigenstate of the system Hamiltonian, namely $|{\vec{k}_{\alpha}}\rangle  = a^\dagger _{\vec{k}_\alpha}|0\rangle $ and $|\vec{k}_{\beta}\rangle  = b^\dagger _{\vec{k}_{\beta}}|0\rangle $, with $(H - \varepsilon_{\vec{k}_{\gamma}})|\vec{k}_\gamma\rangle  = 0$ for $\gamma = \alpha$ or $\beta$.
 
The creation operators of the one-pair eigenstates $|{i}\rangle = B^\dagger _i|0\rangle$ of the system Hamiltonian, with $(H-E_i)|{i}\rangle  = 0$, can be written in terms of these free fermion operators as
\begin{equation} 
B^\dagger _i = \sum{a^\dagger _{\vec{k}_\alpha}b^\dagger _{\vec{k}_\beta}\langle \vec{k}_{\beta}\vec{k}_{\alpha}|{i}\rangle }.
\label{3-3}
\end{equation}
In the same way, the fermion pair creation operators can be written in terms of the $B^\dagger_i$'s as
\begin{equation} 
a^\dagger _{\vec{k}_{\alpha}}b^\dagger _{\vec{k}_{\beta}} = \sum{B^\dagger _i \langle i|\vec{k}_{\alpha}\vec{k}_{\beta}\rangle },
\label{3-4}
\end{equation}
 $\langle \vec{k}_{\beta}\vec{k}_{\alpha}|i\rangle $ is the wave function of the eigenstate $i$ in momentum space if $(H_{\alpha}, H_{\beta})$ only contain free energy contributions. These two equations (\ref{3-3}) and (\ref{3-4}) are quite fundamental in the composite boson many-body theory as they allow us to ``open" the composite bosons into their fermionic components to let the fermions interact through their exact interaction potentials and then to ``close"  these fermions back into composite bosons after their interactions.
 
\begin{figure}
 \begin{center}
 \includegraphics[width=0.6\textwidth]{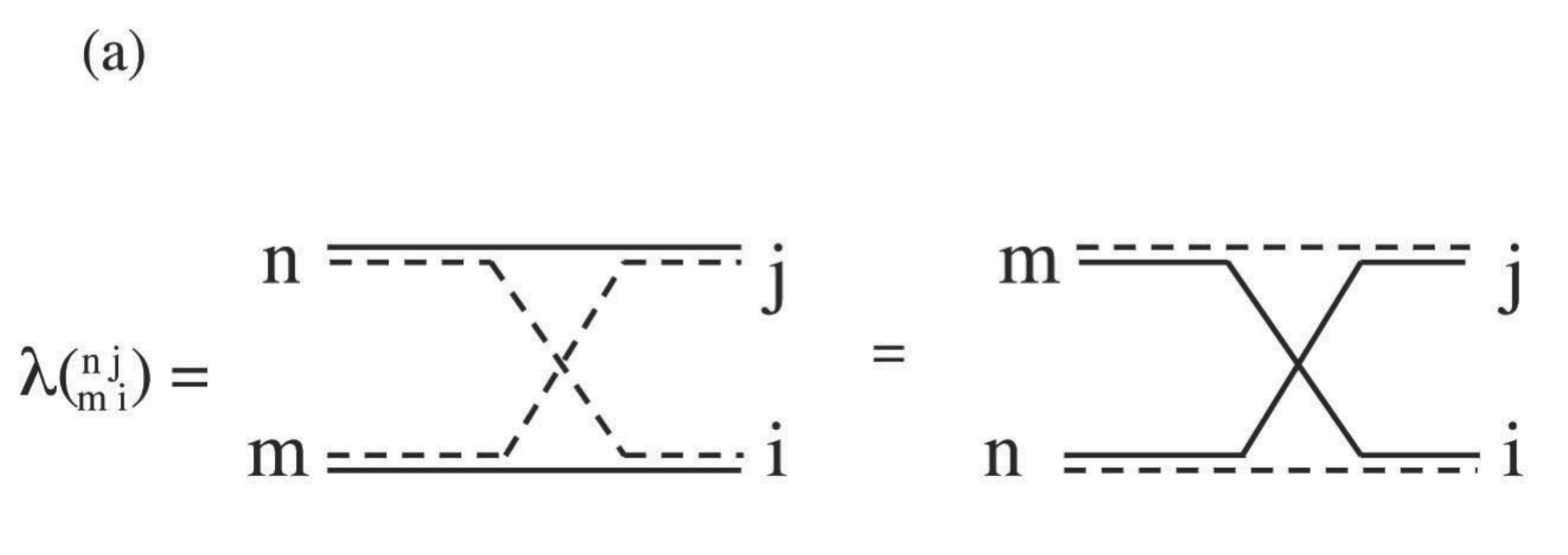}
 
 \includegraphics[width=0.33\textwidth]{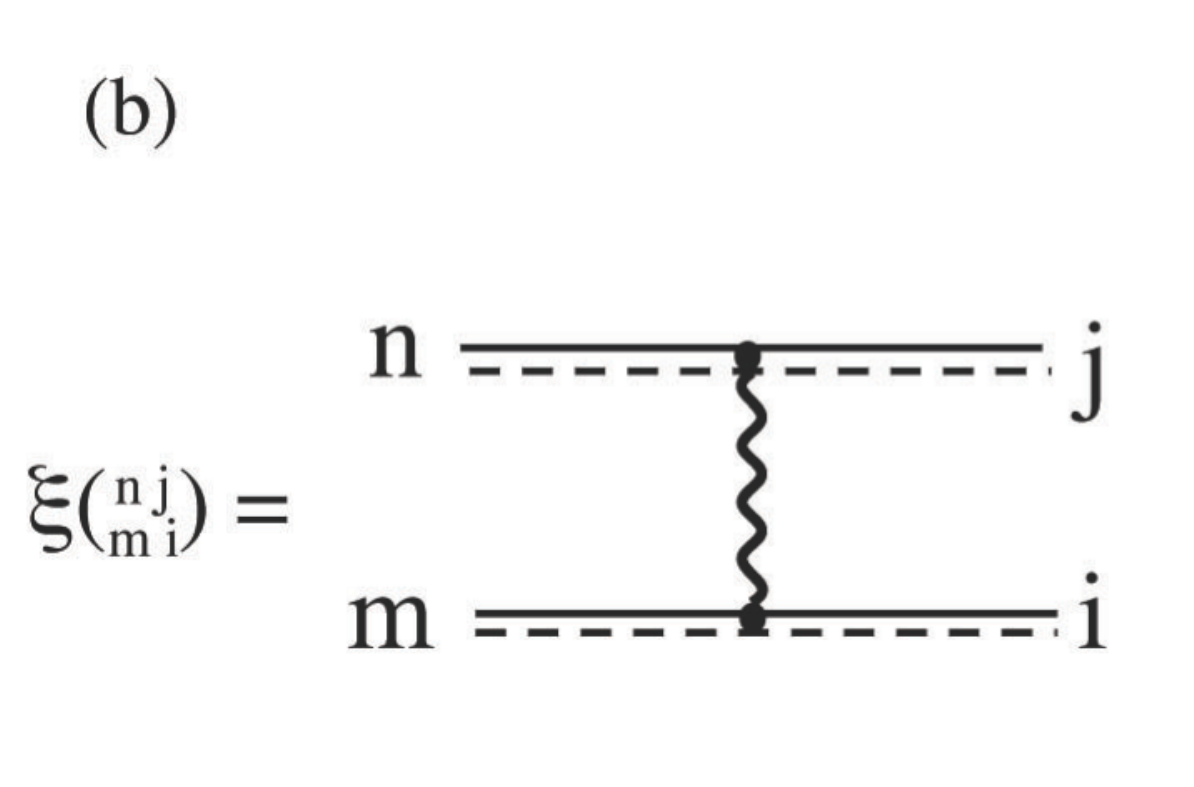}

 \includegraphics[width=0.43\textwidth]{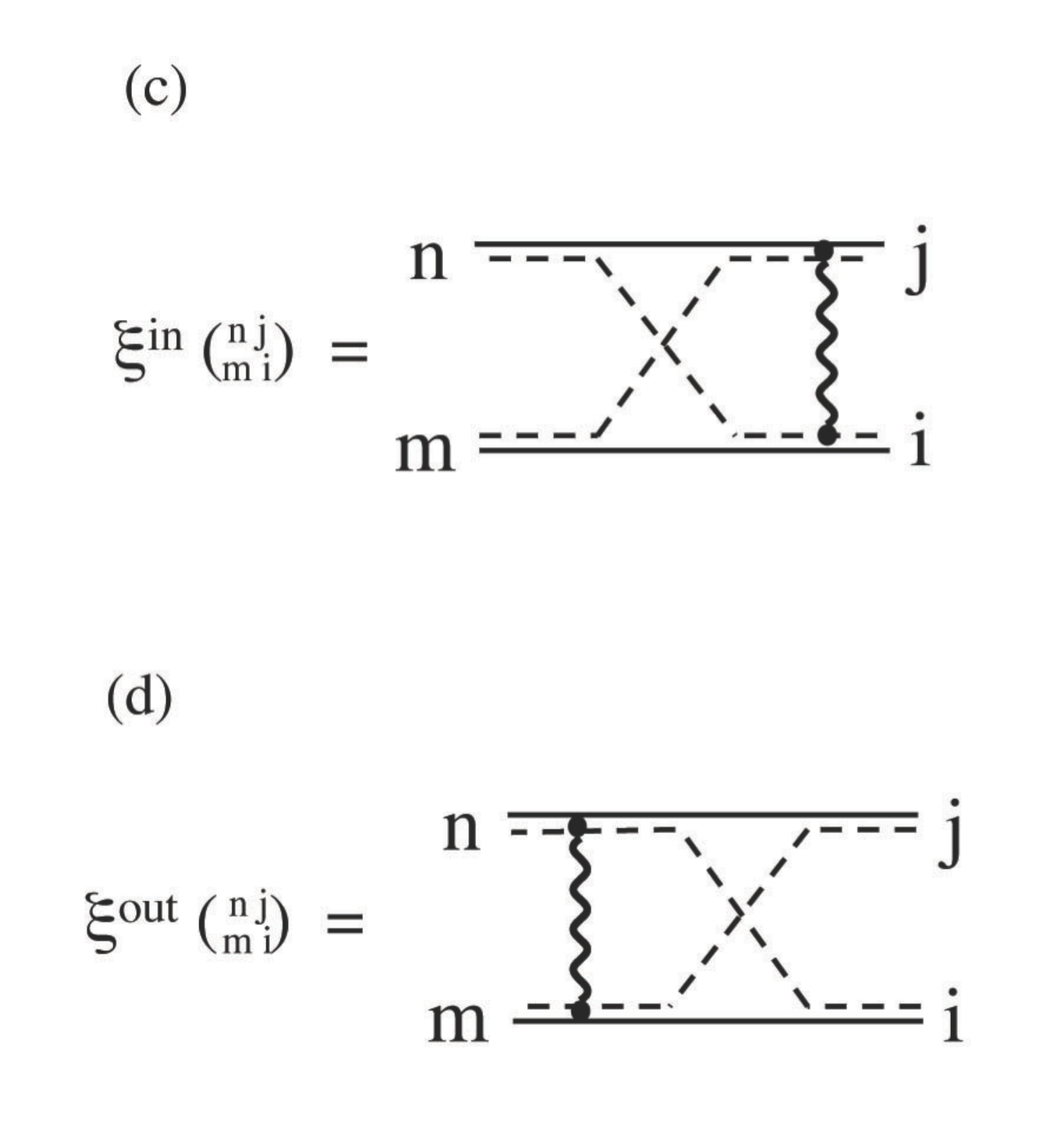}
 \end{center}
 \caption{a) Pauli scattering $\lambda(^n_m \left.\right.^j_i)$, as defined in equation (\ref{cbD}), for fermion exchange between two composite bosons in the absence of fermion interaction. Note that this scattering can be seen as an exchange of fermions $\beta$ or an exchange of fermions $\alpha$ with the indices $(m,n)$ permutated. b) Interaction scattering $\xi(^n_m \left.\right.^j_i)$, as defined in equation (\ref{cbV}), for fermion interaction between to composite bosons in the absence of fermion exchange. c) Exchange interaction scattering $\xi^{in}(^n_m \left.\right.^j_i)$, as defined in equation (\ref{inoutdef}), when the fermion exchange takes place {\em after} the interaction, so that these interactions are between the ``in'' cobosons $(i,j)$. d) Exchange interaction scattering  $\xi^{out}(^n_m \left.\right.^j_i)$, as defined in equation (\ref{inoutdef}), when the fermion exchange occurs {\em before} the interaction, so that these interactions are between the ``out'' cobosons $(m,n)$.}
 \label{fig2}
\end{figure}

The $B^\dagger _i$'s are composite boson operators. Indeed, their behavior is like bosons, since
\begin{eqnarray} 
&\mbox{$[$}B_m,B_i] = 0.
\label{cbB}
\end{eqnarray}
However, since these operators are such that
\begin{eqnarray}
&\mbox{$[$}B_m,B^\dagger _i] = \delta_{mi} - D_{mi},
\label{cbB'}
\end{eqnarray} 
they differ from elementary bosons, the deviation from elementary boson statistics appearing in the operator $D_{mi}$.  This operator is such that $D_{mi}|0\rangle = 0$, as can be seen by taking the above equation acting on the vacuum state $|0\rangle$. The ``Pauli scatterings,'' which describe fermion exchanges between two cobosons in the absence of fermion interaction, are obtained from these deviation-from-boson operators $D_{mi}$, through the commutation relation
\begin{equation} 
[D_{mi},B^\dagger _j] = \sum{\left\{\lambda{ (^n_m \left. \right.^j_i  )}+(m\leftrightarrow n)\right\}B^\dagger _n},
\label{cbD}
\end{equation}
where the second term means the same as the first but with $m$ and $n$ interchanged. $\lambda{(^n_m \left. \right.^j_i  )}$, which is a dimensionless factor, describes the fermion exchange between cobosons $(i,j)$ shown in Fig.~\ref{fig2}(a).

To get interaction scatterings which are energylike quantities, we introduce the ``creation potential'' $V^\dagger _i$, which arises from the commutation relation \cite{r14,r23}
\begin{equation}
[H,B^\dagger _i] = E_iB^\dagger _i + V^\dagger _i,
\label{cbH}
\end{equation}
so that $V^\dagger _i|0\rangle  = 0$ for $B^\dagger _i|0\rangle $ being the $H$ eigenstate with energy $E_i$ . The interaction scatterings $\xi{(^n_m \left. \right.^j_i  )}$ then appear through the commutation relation \cite{r14}
 \begin{equation}
[V^\dagger _i,B^\dagger _j] = \sum{\xi {(^n_m \left. \right.^j_i  )}B^\dagger _mB^\dagger _n}.
\label{cbV}
\end{equation}
The prefactor  $\xi{ (^n_m \left. \right.^j_i  )}$, shown in Fig.~\ref{fig2}(b), describes the direct scattering of the ``incoming'' cobosons $(i,j)$ resulting from the interactions of their fermions, the  ``outgoing'' cobosons $(m,n)$ being made with the same fermion pairs.  From $\xi{ (^n_m \left. \right.^j_i  )}$, it is possible to construct the ``in'' and ``out'' exchange interaction scatterings shown in Figs.~\ref{fig2}(c) and \ref{fig2}(d) through a sequence of direct interaction scattering and dimensionless Pauli scattering, according to
 \begin{eqnarray}
\xi^{in}{ (^n_m \left. \right.^j_i  )}   & =& \sum \lambda{  (^n_m \left. \right.^q_p  )} \xi{  (^q_p \left. \right.^j_i  )}\nonumber \\
\xi^{out}{ (^n_m \left. \right.^j_i  )} &=& \sum \xi{  (^n_m \left. \right.^q_p  )} \lambda{ (^q_p \left. \right.^j_i  )}
\label{inoutdef}
\end{eqnarray} 
 In these exchange interaction scatterings, the ``in'' and ``out'' cobosons are made with different pairs, the fermion interactions taking place between the ``in'' cobosons in $\xi^{in}{ (^n_m \left. \right.^j_i  )}$ and between the ``out'' cobosons in $\xi^{out}{(^n_m \left. \right.^j_i  )}$.  These exchange interaction scatterings are linked to the Pauli scattering through \cite{r15}
\begin{equation} 
\xi^{in}{(^n_m \left. \right.^j_i  )} - \xi^{out}{ (^n_m \left. \right.^j_i  )} = (E_m+E_n-E_i-E_j)\lambda{ (^n_m \left. \right.^j_i  )}.
\label{pscatt}
\end{equation}
 so that they are equal for energy conserving processes. Since both, interaction processes and fermion exchanges conserve momenta, all these scatterings differ from zero only for $\vec{Q}_m+\vec{Q}_n = \vec{Q}_i+\vec{Q}_j$.
 
In the following, we will see that the physically relevant combinations of energylike scatterings appear to be
\begin{equation}
\hat{\xi}{ (^n_m \left. \right.^j_i  )} = \xi{ (^n_m \left. \right.^j_i  )} - \xi^{in}{ (^n_m \left. \right.^j_i  )}
\label{I14}
\end{equation}
that we are going to symmeterize as
\begin{equation} 
\hat{\xi}_{mn;ij} = \frac{1}{2}\hat{\xi}{ (^n_m \left. \right.^j_i  )} + (m\leftrightarrow n),
\label{symmxi}
\end{equation}
in order to have $\hat{\xi}_{mn;ij} = \hat{\xi}_{nm;ij} = \hat{\xi}_{mn;ji}$.  In the same way, we introduce the symmetrized Pauli scatterings $\lambda_{mn;ij}$, equal to  $[\lambda (^n_m \left. \right.^j_i  ) + (m \leftrightarrow n)]/2$.

\subsection{Many-body effects}

To derive many-body effects with identical cobosons, it is convenient to iterate the four commutation relations (\ref{cbB})-(\ref{cbV}) on which the composite boson many-body theory is based. This leads to \cite{r14,r15}
\begin{eqnarray}
[B_m,B^{\dagger N}_o] &=& NB^{\dagger  N-1}_o (\delta_{mo} - D_{mo}) - N(N-1)B^{\dagger  N-2}_o\sum \lambda{(^n_m \left. \right.^o_o  )} B^\dagger_n \label{mcbB}
\end{eqnarray}
\begin{eqnarray}
\mbox{$[$}D_{mi},  B^{\dagger N}_o] &=& NB^{\dagger  N-1}_o \sum{\left\{ \lambda{(^n_m \left. \right.^i_o  )}+ (m\leftrightarrow n)\right\} }B^\dagger _n \label{mcbD}
\end{eqnarray}
for many-body effects dealing with fermion exchanges and
\begin{eqnarray}
\mbox{$[$}H,B^{\dagger N}_o] &=& NB^{\dagger  N-1}_o(E_oB^\dagger _o + V^\dagger _o) + \frac{N(N-1)}{2}B^{\dagger  N-2}_o\sum{\xi{(^n_m \left. \right.^o_o  )}B^\dagger _mB^\dagger _n} \label{mcbH}\\
\mbox{$[$}V^\dagger _i,B^{\dagger N}_o] &=& NB^{\dagger  N-1}_o\sum{\xi{(^n_m \left. \right.^o_i  )}}B^\dagger _mB^\dagger _n
\label{mcbV}
\end{eqnarray}
for many-body effects dealing with fermion interactions.
The $N$ prefactors in front of the $\lambda$ and $\xi$ terms come from the number of ways among $N$ to choose the cobosons in state $o$ involved in these scatterings.  This gives $N$ when only one coboson in state $o$ is involved and ${N(N-1)}/{2}$ when two cobosons in state $o$ are involved, the additional 2 in (\ref{mcbB}) coming from the fact that these cobosons can exchange either their fermion $\beta$ or their fermion $\alpha$, the ($\alpha\leftrightarrow\beta$) exchange being identical to a $(i\leftrightarrow j)$ exchange, as seen in Fig.~\ref{fig2}(a).

In typical problems dealing with $N$ cobosons, most of them are in the same state. Scalar products of such $N$ coboson states are easy to expand \cite{r17} in terms of 
\begin{equation}
\langle 0 | B_o^N B_o^{\dagger N}| 0\rangle = N!F_N. 
\end{equation}
 The normalization factor $F_N$ reduces to 1 for the case of elementary bosons, but it is not a number of the order of 1 for cobosons due to the Pauli exclusion principle between the fermionic components of the particles. Indeed, from the recursion relation obeyed by $F_N$, it has been shown \cite{r15} that, for large samples, $F_N$ is exponentially small ($F_N \sim e^{-N\eta}$), where $\eta$ is the dimensionless parameter associated with density defined by (\ref{eta}).  However, in calculations of physical quantities, $F_N$ always appears in ratios $F_{N-n}/F_N$ which, for $n$ small, is equal to 1 within corrections of the order of $\eta$ 
\beq
\frac{F_{N-1}}{F_N} = 1 + {\cal O}(\eta).
\eeq

\subsection{Coherent superposition of cobosons}
\label{sect.cohbos}

Besides many-body states with a given number of particles in a specific eigenstate, we wish also to consider coherent superpositions of cobosons, i.e., states of the form
\begin{equation}
B^{\dagger} = a'B^{\dagger}_{o'} + a''B^{\dagger}_{o''}.
\end{equation}
While calculations for $N = 2$ are easy to perform by simply expanding $(a'B^\dagger _{o'} + a''B^\dagger _{o''})^2$, calculations for large $N$ are more tricky.  For those, we are forced to keep the coherent state $B^\dagger $ as an entity, and to produce equations similar to (\ref{mcbB})-(\ref{mcbV})  with $B^{\dagger}_o$ replaced by  $B^\dagger$. 

\subsubsection{Fermion exchanges}
To get the fermion exchanges in a convenient way, let us first introduce the operator $D_m$ through the commutator
$[B_m,B^\dagger ] =\delta_m- D_m$
where $\delta_m$ is defined in (\ref{2-7}). $D_m$ reads in terms of the deviation-from-boson operators as $D_m = a'D_{mo'} + a''D_{mo''}$.
From it, we then construct the $\lambda_{mn}$'s through the commutator
$[D_m,B^\dagger ] = 2\sum{\lambda_{mn}B^\dagger _n}$.
They read in terms of the Pauli scatterings as
\begin{equation}
\lambda_{mn} = a'^{2}\lambda{(^n_m \left. \right.^{o'}_{o'}  )} + a''^2\lambda{(^n_m \left. \right.^{o''}_{o''} )}
 + a'a''\left[\lambda{(^n_m \left. \right.^{o''}_{o'}  )} + \lambda{(^n_m \left. \right.^{o'}_{o''}  )}\right].
\label{gcblam}
\end{equation}
Iteration of these two commutators allows us to write $[D_m,B^{\dagger N}]$ and $[B_m,B^{\dagger N}]$ in terms of this $\lambda_{mn}$.We find 
\begin{equation}
[D_m,B^{\dagger N}] \ = \  2NB^{\dagger  N-1}\sum{\lambda_{mn}B^\dagger _n}
\label{Dmult}
\end{equation}
\begin{equation}
[B_m,B^{\dagger N}] = NB^{\dagger  N-1}(\delta_m - D_m) - N(N-1)B^{\dagger  N-2}\sum{\lambda_{mn}B^\dagger _n}
\label{Bmult}
\end{equation}
From these two commutators, it is also possible to show that 
$[B,B^\dagger ] = 1 - D $ while
$[D,B^\dagger ] = 2\sum{\lambda_nB^\dagger _n}$
in which the prefactor $\lambda_n$ reads in terms of the Pauli scatterings as
\begin{eqnarray}
\lambda_n = a'|a'|^2\lambda{(^n_{o'} \left. \right.^{o'}_{o'}  )}
 + a''|a''|^2\lambda{(^n_{o''} \left. \right.^{o''}_{o''}  )}\nonumber\\
+ a''|a'|^2\left\{\lambda{(^n_{o'} \left. \right.^{o''}_{o'}  )} + \lambda{(^n_{o'} \left. \right.^{o'}_{o''}  )} \right\} 
 + a'|a''|^2\left\{ \lambda{(^n_{o''} \left. \right.^{o'}_{o''}  )} + \lambda{(^n_{o''} \left. \right.^{o''}_{o'}  )}\right\}.
\end{eqnarray}
Similar commutators for $N$, obtained by iteration, read in terms of this $\lambda_n $ as
\begin{eqnarray}
[D,B^{\dagger N}] &=& 2NB^{\dagger  N-1}\sum{\lambda_nB^\dagger _n}\\
\mbox{$[$}B,B^{\dagger N}] &=& NB^{\dagger  N-1}(1-D) - N(N-1)B^{\dagger  N-2}\sum{\lambda_nB^\dagger _n}.
\label{gmcbD}
\end{eqnarray}
These two commutation relations, along with equations (\ref{Dmult},\ref{Bmult}), are the equivalent of equations (\ref{mcbB}) and (\ref{mcbD}) for coherent superpositions of cobosons. 

\subsubsection{Fermion interactions}

To perform calculations for $N$ of these coherent superpositions in a convenient way, we also need similar commutators for the interaction part.  They are obtained through
$[H,B^\dagger ] = E_o\tilde{B}^\dagger  + V^\dagger$ ,
in which we have set $V^\dagger  = a'V^\dagger _{o'} + a''B^\dagger _{o''}$ while  \begin{equation}
\tilde{B}^\dagger  = {(a'E_{o'}B^\dagger _{o'} + a''E_{o''}B^\dagger _{o''})}/{E_o} \end{equation}
So that $\tilde{B}^{\dagger}$ reduces to $B^\dagger $ for $E_{o'} \approx E_{o''}\approx{E_o}$.  We are then led to define $\xi_{mn}$ through $[V^\dagger ,B^\dagger ] = \sum{\xi_{mn}B^\dagger _mB^\dagger _n}$. This scalar  is found to be
\begin{equation}
%(V 16) 
\xi_{mn} = a'^2\xi{(^n_{m} \left. \right.^{o'}_{o'}  )} + a''^2\xi{(^n_{m} \left. \right.^{o''}_{o''}  )} + a'a''\left\{\xi{(^n_{m} \left. \right.^{o''}_{o'}  )} +\xi{(^n_{m} \left. \right.^{o'}_{o''}  )}\right\},
\label{V16}
\end{equation}
so that the iteration of these commutators leads to
\begin{eqnarray}
%(V 17)
 [V^\dagger ,B^{\dagger N}] &=& NB^{\dagger  N-1}\sum{\xi_{mn}B^\dagger _mB^\dagger _n}\\
%(V 18)  
\mbox{$[$}H,B^{\dagger N}] &=& NB^{\dagger  N-1}(E_o\tilde{B}^\dagger  + V^\dagger ) + \frac{N(N-1)}{2}B^{\dagger N-2}\sum{\xi_{mn}B^\dagger _mB^\dagger _n}.
\label{V18}
\end{eqnarray}

\subsubsection{Normalization factors}

As for single composite bosons, the norm of the state made of $N$ identical coherent superpositions of cobosons is going to play a key role in the many-body physics of these systems.  This leads us to introduce, just as we defined $F_N$ for single cobosons, the normalization factor $G_N$, determined by
\begin{equation}
%(V 19) 
\langle 0|B^NB^{\dagger N}|0\rangle  = N!G_N.
\end{equation}
$G_N$ is expected to be exponentially small due to the many exchanges which take place between the fermions of the $N$ coherent bosons.  As for $F_N$, we reach $G_N$ through the recursion relation it obeys.  From (\ref{gmcbD}) and the fact that $D|0\rangle  = 0$, we find
\begin{eqnarray}
%(V 20)  
\langle 0|B^NB^{\dagger N}|0\rangle  &=& \langle 0|B^{N-1}[B,B^{\dagger N}]|0\rangle \nonumber\\
& =& N(N-1)!G_{N-1} - N(N-1)\sum{\lambda_n\langle 0|B^{N-1}B^\dagger _nB^{\dagger  N-2}|0\rangle }
\end{eqnarray}
We now use (\ref{Bmult}), to get $\langle 0|B^{N-1}B^\dagger _n$.  This leads to
\begin{equation}
%(V 21) 
G_N = G_{N-1} - (N-1)\lambda ^{(1)}G_{N-2} + (N-1)(N-2)\lambda^{(2)}G_{N-3} + \ldots,
\label{GN}
\end{equation}
where the precise value of the first order term $\lambda^{(1)}$ is 
\begin{eqnarray}
%(V 22) 
\lambda^{(1)} &=& \sum{\lambda_n\delta_n^*} = |a'|^4\lambda{(^{o'}_{o'} \left.  \right.^{o'}_{o'}  )} + |a''|^4\lambda{(^{o''}_{o''} \left. \right.^{o''}_{o''}  )}
 + 2|a'a''|^2\left\{\lambda{(^{o''}_{o'}\left. \right.^{o''}_{o'}  )} + \lambda{(^{o''}_{o'} \left. \right.^{o'}_{o''}  )} \right\}
 \label{lam2}
 \end{eqnarray}
Equations (\ref{GN}) and (\ref{lam2}) show that like $F_N$, the $G_N$ correction to the bare elementary boson normalization factor $N!$, although not of the order of 1, is such that
\begin{equation}
\frac{G_{N-1}}{G_N} \simeq 1 + N\lambda^{(1)} = 1 + {\cal O}(\eta)
 \label{lam3}
\end{equation}

\subsection{Mixture of cobosons}
\begin{figure}
\begin{center}
 \includegraphics[width=0.9\textwidth]{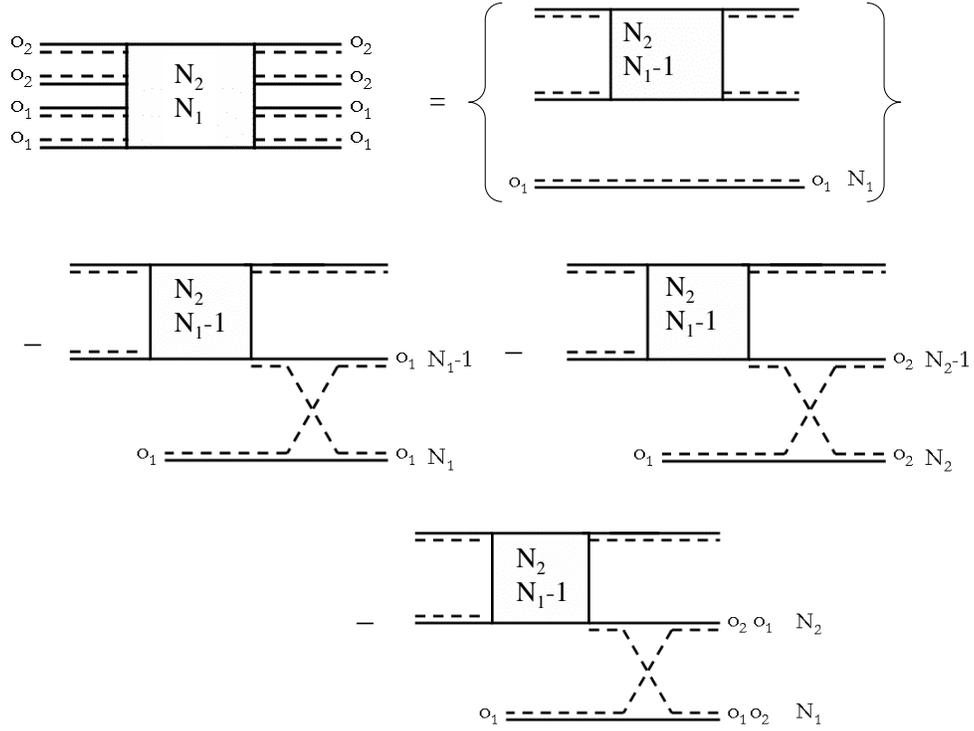}
\end{center}
 \caption{Shiva diagram expansion for the normalization factor $\langle 0 | B^{N_2}_{o_2}B^{N_1}_{o_1}B^{\dagger N_1}_{o_1}B^{\dagger N_2}_{o_2}|0\rangle$ appearing in equation (\ref{V40}).}
 \label{fig3}
\end{figure}

\begin{figure}
  \begin{center}
 \includegraphics[width=0.8\textwidth]{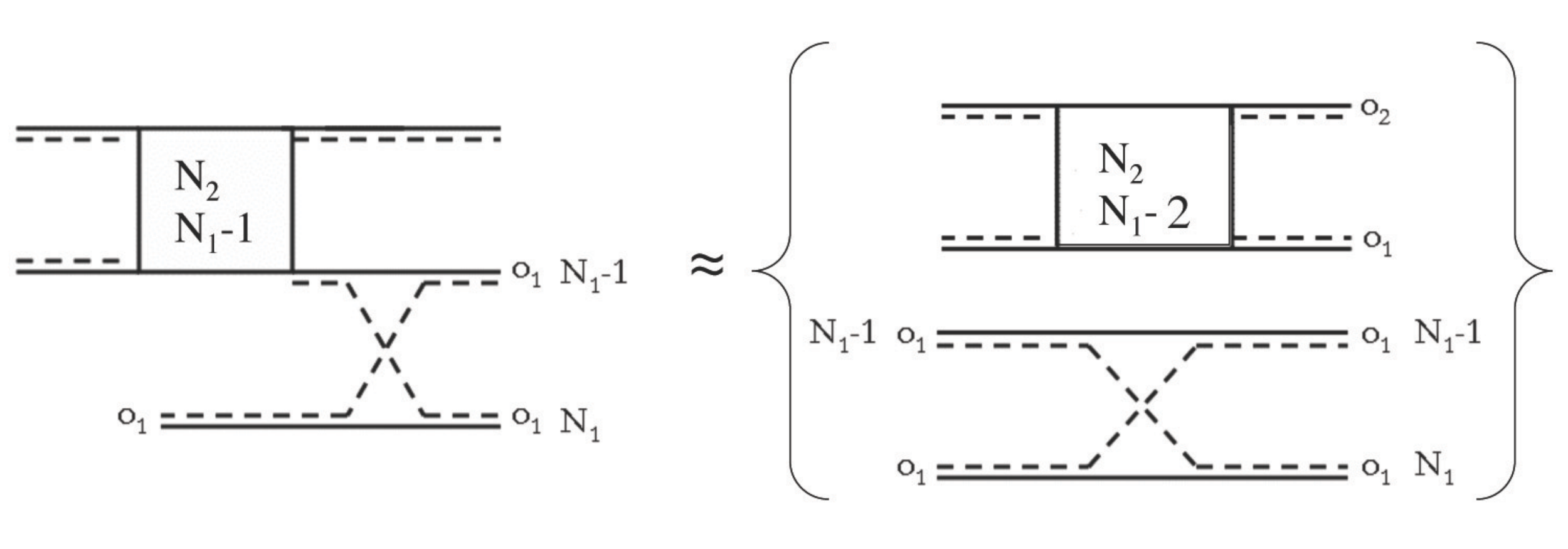}
\end{center}
 \caption{Shiva diagram expansion for the first exchange contribution $N_1(N_1-1)A_{11}$ to $\langle 0 | B^{N_2}_{o_2}B^{N_1}_{o_1}B^{\dagger N_1}_{o_1}B^{\dagger N_2}_{o_2}|0\rangle$, as given in equation (\ref{V41}).}
\label{fig4}
\end{figure}

In this paper, we are also going to consider a mixture of cobosons of the form $B^{\dagger N_1}_{o_1}B^{\dagger N_2}_{o_2}|0\rangle$.  Like $F_N$ for many-body effects between $N$ identical cobosons, $G_{N_1,N_2}$, defined by
\begin{equation}
\langle 0 | B^{N_2}_{o_2}B^{N_1}_{o_1} B^{\dagger N_1}_{o_1}B^{\dagger N_2}_{o_2}|0\rangle = N_1!N_2!G_{N_1,N_2},
\label{g12norm}
\end{equation}
is a  key factor for many-body effects between $N_1$ cobosons in state $o_1$ and $N_2$ cobosons in state $o_2$.   This factor, which would be exactly 1 for elementary bosons, differs from 1 due to the many fermion exchanges which exist not only among the $N_1$ cobosons in state $o_1$ and among the $N_2$ cobosons in state $o_2$, but also between the cobosons in state $o_1$ and the cobosons in state $o_2$.  They make $G_{N_1,N_2}$ not of the order of 1 but exponentially small.  As with $F_N$, we can determine $G_{N_1,N_2}$ through the recursion relation it fufills.  To get this recursion relation, we use equation (\ref{mcbB}) to rewrite $B_{o_1}B^{\dagger N_1}_{o_1}$.  This leads to
\begin{eqnarray}
%(V 38) 
\langle 0|B^{N_2}_{o_2}B^{N_1}_{o_1}B^{\dagger N_1}_{o_1}B^{\dagger N_2}_{o_2}|0\rangle  &=& 
\langle 0|B^{N_2}_{o_2}B^{N_1-1}_{o_1} \left\{ B^{\dagger N_1}_{o_1}B_{o_1} + N_1B^{\dagger N_1-1}_{o_1}(1 - D_{o_1o_1})\nonumber \right. \\ 
&& \left. - N_1(N_1 - 1)B^{\dagger N_1-2}_{o_1}\sum{\lambda}{(^{i}_{o_1}\left. \right.^{o_1}_{o_1}  )} B^\dagger _i \right\} B^{\dagger N_2}_{o_2}|0\rangle 
\label{2opnorm}
\end{eqnarray}
The 1 in the second term in the bracket readily gives
$N_1\left[(N_1-1)!N_2!G_{N_1-1,N_2}\right]$.
The first term in the bracket is calculated using (\ref{mcbB}) for $B_{o_1}B^{\dagger N_2}_{o_2}|0\rangle $, while the third term is calculated using (\ref{mcbD}) for $D_{o_1o_1}B^{\dagger N_2}_{o_2}|0\rangle $.  This allows us to write the norm of $B^{\dagger N_1}_{o_1}B^{\dagger N_2}_{o_2}|0\rangle $ given in equation (\ref{g12norm}) as
\begin{equation}
%(V 40)  
N_2!N_2!G_{N_1,N_2} = N_1[(N_1-1)!N_2!G_{N_1-1,N_2}] - N_1(N_1-1)A_{11}
-N_2(N_2-1)A_{22} - N_1N_2A_{12},
\label{V40}
\end{equation}
where $A_{11},~A_{22},$ and $A_{12}$ contain one Pauli scattering explicitly, so that they are related to fermion exchanges.  Equation (\ref{V40}) is shown in Fig.~\ref{fig3}, the factors $(N_1,N_2)$ coming from the number of ways we can choose the cobosons involved in fermion exchanges with the coboson $o_1$ on the left.  The leading term for $N_1(N_1-1)A_{11}$, shown in Fig.~\ref{fig4}, reduces to
\begin{equation}
\label{V41} 
(N_1-1)\lambda{(^{o_1}_{o_1}\left. \right.^{o_1}_{o_1}  )}N_1(N_1-1)\biggl[(N_1-2)!N_2!G_{N_1-2,N_2}\biggr].
\end{equation}

\begin{figure}
\begin{center}
\includegraphics[width=0.8\textwidth]{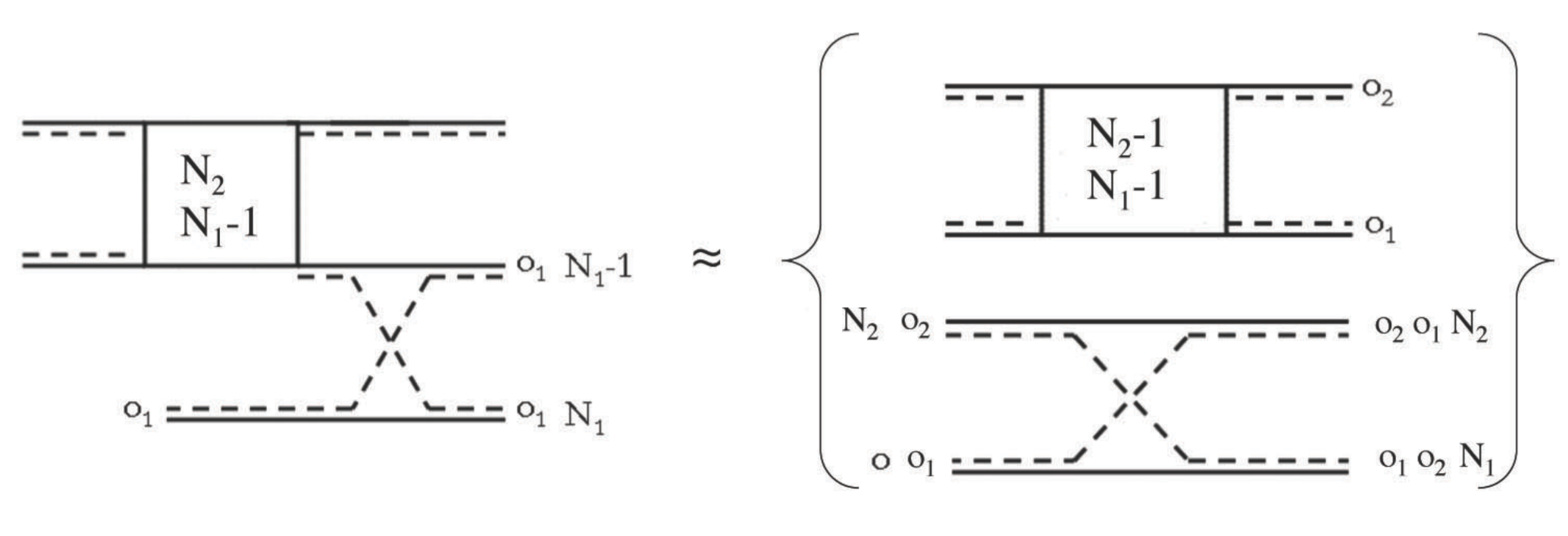}
\end{center}
\caption{Same as Figure 4 for the third exchange contribution $N_1N_2A_{12}$, as given in equation (\ref{V42}).}
\label{fig5}
\end{figure}

Similarly, the leading term for $N_1N_2A_{12}$, shown in Fig.~\ref{fig5}, is
\begin{equation}
\label{V42}
N_2\left[\lambda{(^{o_2}_{o_1}\left. \right.^{o_2}_{o_1}  )} + \lambda{(^{o_2}_{o_1}\left. \right.^{o_1}_{o_2}  )}\right]N_1N_2\biggl[(N_1-1)!(N_2-1)!G_{N_1-1,N_2-1}\biggr].
\end{equation}
They contain fermion exchanges between two cobosons. On the opposite, the leading term of $N_2(N_2-1)A_{22}$ must have fermion exchange between 3 cobosons since $\lambda{(^{o_1}_{o_2}\left. \right.^{o_2}_{o_2}  )} = 0$ for $o_1 \neq o_2$.  Therefore, we end with
\begin{eqnarray}
%(V 43)  
G_{N_1,N_2} = G_{N_1-1,N_2} - (N_1-1)\lambda{(^{o_1}_{o_1}\left. \right.^{o_1}_{o_1}  )}
G_{N_1-2,N_2}
- N_2\left(\lambda{(^{o_2}_{o_1}\left. \right.^{o_2}_{o_1}  )} + \lambda{(^{o_2}_{o_1}\left. \right.^{o_1}_{o_2}  )}\right)G_{N_1-1,N_2-1} + \ldots\nonumber\\
\end{eqnarray}
This shows that, in the same way that $F_{N-1}/F_N \approx 1 + \cal{O}(\eta)$, we have
\begin{equation}
%(V 44)
\frac{G_{N_1-1,N_2}}{G_{N_1,N_2}} = 1 + {\cal O}(\eta_{1}) + {\cal O}(\eta_{2}), 
\end{equation}
so that, even if $G_{N_1,N_2}$ is exponentially small due to the large number of fermion exchanges, the effect of these exchanges on $G_{N_1,N_2}$ ratios are negligible at lowest order in density.

We now have all the tools to tackle many-body effects with a large number of identical cobosons, a large number of coherent cobosons  and a mixture of two large numbers of cobosons. However, since the calculations for large $N$'s are obviously quite technical, in this paper we have chosen to start with $N=2$, as most of the many-body physics can usually be understood from this limit. We are also going to first perform these calculations for elementary bosons, in order to enlighten the differences between elementary and composite particles.

\section{Two elementary bosons}
\label{sect.two1}
\setcounter{equation}{0}

\subsection{Single state}

Let us start with a state having two identical elementary bosons, $|\bar{\phi}_0\rangle  = \bar{B}^{\dagger 2}_o|0\rangle $.  Since $\bar{B}_i|\bar{\phi}_0\rangle  = 2\delta_{io}\bar{B}^\dagger _o|0\rangle $ due to equation (\ref{mebB}), we find that $\langle \bar{\phi}_0|\bar{\phi}_0\rangle  = 2$ while
\begin{equation}
\langle \bar{\phi}_0| \bar{H_0}|\bar{\phi}_0\rangle = \sum E_i \{\langle \bar{\phi}_0|\bar{B}^{\dagger}_i \}\{\bar{B}_i|\bar{\phi}_0\rangle \} = 4E_o.  
\end{equation}
\begin{equation}
%(II 1)  
\langle \bar{\phi}_0|\bar{V}|\bar{\phi}_0\rangle  = \frac{1}{2}\sum{\bar{\xi}{ (^{n}_{m}\left. \right.^{j}_{i}  )}{\langle \bar{\phi}_0|\bar{B}^\dagger _m\bar{B}^\dagger _n}{\bar{B}_j\bar{B}_i|\bar{\phi}_0\rangle }} = 2\bar{\xi}{ (^{o}_{o}\left. \right.^{o}_{o}  )}.
\end{equation}
This readily gives the Hamiltonian mean value in this two-elementary boson state as
\begin{equation}
%(II 3)  
\frac{\langle \bar{\phi}_0|\bar{H}|\bar{\phi}_0\rangle }{\langle \bar{\phi}_0|\bar{\phi}_0\rangle } = 2E_o + \bar{\xi}{  (^{o}_{o}\left. \right.^{o}_{o}  )} = 2E_o + \bar{\xi}_{oo;oo},
\label{II3}
\end{equation}
where $\bar{\xi}_{mn;ij}$ is defined in terms of $\bar{\xi}{  (^{n}_{m}\left. \right.^{i}_{j}  )} $ as in (\ref{symmxi}).

\subsection{Fragmented state}

We now consider $|\bar{\phi}_{12}\rangle  = \bar{B}^\dagger _{o_1}\bar{B}^\dagger _{o_2}|0\rangle $ for elementary bosons in states $o_1\neq o_2$.  
From $\bar{B}_i\bar{\phi}{12}\rangle $, we readily find $\langle \bar{\phi}_{12}|\bar{\phi}_{12}\rangle  = 1$, while $\langle \bar{\phi}_{12}|\bar{H}_0|\bar{\phi}_{12}\rangle  = E_{o_1} + E_{o_2}$.  
In the same way, $\bar{B}_j\bar{B}_i|\bar{\phi}_{12}\rangle  = (\delta_{jo_2}\delta_{io_1} + \delta_{jo_1}\delta_{io_2})|0\rangle $ leads to 
\begin{equation}
%(II 4)  
\langle \bar{\phi}_{12}|\bar{V}|\bar{\phi}_{12}\rangle  = \bar{\xi}{  (^{o_2}_{o_1}\left. \right.^{o_2}_{o_1}  )} +\bar{\xi}{ (^{o_1}_{o_2}\left. \right.^{o_2}_{o_1}  )},
\end{equation}
so that we end up with
\begin{equation}
%(II 5)  
\frac{\langle \bar{\phi}_{12}|\bar{H}|\bar{\phi}_{12}\rangle }{\langle \bar{\phi}_{12}|\bar{\phi}_{12}\rangle } = E_{o_1} + E_{o_2} + 2\bar{\xi}_{o_1o_2;o_1o_2} \approx 2E_o + 2\bar{\xi}_{oo;oo}
\label{II5}
\end{equation}
for $o_1\approx o_2\approx o$.

\subsection{Coherent superposition}

The third state of interest is the coherent superposition of states, $|\bar{\phi}\rangle  = \bar{B}^{\dagger 2}|0\rangle $, with $\bar{B}^\dagger $ defined as in (3-20), with $B_{o'}^\dagger $ replaced by $\bar{B_{o'}}^\dagger$. The prefactors $(a',a'')$ are chosen such that $|a'|^2 + |a''|^2 = 1$  in order for the state to be normalized. We also impose that the $(o',o'')$ states differ at least through their center of momenta $(\vec{Q}_{o'} \neq \vec{Q}_{o''})$  in order for scatterings like $\bar{\xi}{  (^{o''}_{o'}\left. \right.^{o'}_{o'}  )}$ to cancel due to momentum conservation. From equation (\ref{cohelmany}) taken for $N=2$, we readily find
$\langle \bar{\phi}|\bar{\phi}\rangle  = 2$ while $\langle \bar{\phi}|\bar{H}_0|\bar{\phi}\rangle  = 4(|a'| ^2E_{o'} + |a''| ^2E_{o''})$.  In the same way, from 
$\bar{B}_j \bar{B}_i|\bar{\phi}\rangle  = 2(a'\delta_{io'} + a''\delta_{io''}) (a'\delta_{jo'} + a''\delta_{jo''})|0\rangle $ 
and the fact that  the nonzero $\bar{\xi}$ reduce for $\vec{Q}_{o'} \ne \vec{Q}_{o''}$ to the diagonal terms $\bar{\xi}{ (^{i}_{j}\left. \right.^{i}_{j}  )}$, with $(i,j) = (o'~ \mbox{or}~ o'')$, or to the cross term $\bar{\xi}{ (^{o'}_{o''}\left. \right.^{o''}_{o'}  )}$,  we find that
\begin{eqnarray}
%(II 7) 
\langle \bar{\phi}|\bar{V}|\bar{\phi}\rangle  = 2|a'| ^4\bar{\xi}{ (^{o'}_{o'}\left. \right.^{o'}_{o'}  )} + 2|a''| ^4 \xi{  (^{o''}_{o''}\left. \right.^{o''}_{o''}  )}  + 4 |a'a''| ^2\left[\xi{ (^{o''}_{o'}\left. \right.^{o''}_{o'}  )} + \xi{  (^{o'}_{o''}\left. \right.^{o''}_{o'}  )}\right].
\end{eqnarray}
Consequently, the Hamiltonian mean value for two elementary bosons in a coherent state reads as
\begin{eqnarray}
%(II 8) 
\frac{\langle \bar{\phi}|\bar{H}|\bar{\phi}\rangle }{\langle \bar{\phi}|\bar{\phi}\rangle } &=& 2(|a'|^2E_{{o'}} + |a''|^2E_{o''}) + |a'|^4\bar{\xi}_{o'o';o'o'} + |a''|^4\bar{\xi}_{o''o'';o''o''} + 4|a'a''|^2\bar{\xi}_{o'o'';o'o''}\nonumber\\
&\approx& 2E_o + (1+2|a'a''|^2)\bar{\xi}_{oo;oo}
\label{II8}
\end{eqnarray}
for $o'\approx o''\approx o$.

From equations (\ref{II3},\ref{II5},\ref{II8}), we thus see that, since the interaction scattering $\xi_{oo;oo}$ must be positive to avoid a density collapse, the energy of two elementary bosons is minimum when the particles are in a single eigenstate, and not a mixture or a coherent superposition of two eigenstates.

\section{Two composite bosons}
\label{sect.two2}
\setcounter{equation}{0}

The calculations for composite bosons will need to explictly treat the exchange between the constituent fermions, through the deviation-from-boson operator $D_{mi}$ and the Pauli scattering $\lambda{(^{n}_{m}\left. \right.^{j}_{i}  )}$.

\subsection{Single state}

For $|\phi_0\rangle  = B^{\dagger 2}_o|0\rangle $, we now have, due to equation (\ref{mcbB}),
\begin{equation}
%(III 1)  
B_m|\phi_0\rangle  = 2\delta_{mo}B^\dagger _o|0\rangle  - 2\sum{\lambda{  (^{n}_{m}\left. \right.^{o}_{o}  )}B^\dagger _n|0\rangle },
\end{equation}
since $D_{mi}|0\rangle  = 0$.  This leads to
\begin{equation}
%(III 2) 
\langle \phi_0|\phi_0\rangle  = 2-2\lambda{ (^{o}_{o}\left. \right.^{o}_{o}  )}
\end{equation}
with $\lambda{ (^{o}_{o}\left. \right.^{o}_{o}  )} = ({33\pi}/{2})({a_B}/{L})^3$ for 3D cobosons \cite{r15} with zero center-of-mass momentum and relative motion wavefunction $\langle r|\nu_0\rangle  = e^{{-r}/{a_B}}/{\sqrt{\pi a_B^3}}$; so that this Pauli scattering goes to zero when the sample size $L$ increases.

Using equation (\ref{mcbH}) and the fact that $V^\dagger _0|0\rangle  = 0$, we find
\begin{equation}
%(III 3)  
H|\phi_0\rangle  = 2E_oB^{\dagger 2}_o|0\rangle  - \sum{\xi{(^{n}_{m}\left. \right.^{o}_{o}  )}B^\dagger _mB^\dagger _n|0\rangle }
\label{III3}
\end{equation}
From equations (\ref{cbB'},\ref{cbD}), it is easy to show that the scalar product of two coboson states reads as \cite{r17}
\begin{equation}
%(III 4)  
\langle 0|B_mB_nB^\dagger _iB^\dagger _j|0\rangle  = \left\{ \delta_{mi}\delta_{nj} - \lambda{(^{n}_{m}\left. \right.^{i}_{j}  )} \right\} + \left\{{m\leftrightarrow n}\right\}
\label{III4}
\end{equation}
So that, from the two above equations, we end up with
\begin{equation}
%(III 5) 
\frac{\langle \phi_0|H|\phi_0\rangle }{\langle \phi_0|\phi_0\rangle } 
= 2E_o + \frac{\xi{(^{o}_{o}\left. \right.^{o}_{o}  )} - \xi^{in} {(^{o}_{o}\left. \right.^{o}_{o}  )} }
{1 - \lambda{(^{o}_{o}\left. \right.^{o}_{o}  )} } =  2E_o + \frac{\hat{\xi}_{oo;oo}}{1-\lambda_{oo;oo}}\approx 2E_o + \hat{\xi}_{oo;oo}.
\label{III5}
\end{equation}
since $\lambda_{oo;oo}$ goes to 0 as $({a_B}/{L})^d$ when the sample size increases.

\subsection{Fragmented state}

We now turn to the fragmented state $|\phi_{12}\rangle  = B^\dagger _{o_1}B^\dagger _{o_2}|0\rangle $ for $(o_1,o_2)$ with different center of mass momenta in order for the Pauli and interaction scatterings to cancel due to momentum conservation if the number of $o_1$ cobosons in the ``in'' and ``out'' states are different. Using eq. (\ref{III4}), its norm is found to be
\begin{eqnarray}
%(III 6)  
\langle \phi_{12}|\phi_{12}\rangle  &=& 1-\lambda{(^{o_2}_{o_1}\left. \right.^{o_2}_{o_1}  )} - \lambda{(^{o_1}_{o_2}\left. \right.^{o_2}_{o_1}  )}\\ & =& 1 - 2\lambda_{o_1o_2;o_1o_2},
\label{III6}
\end{eqnarray}
while from equations (\ref{cbH})-(\ref{cbV}), the Hamiltonian mean value in this fragmented state appears as
\begin{equation}
%(III 7)  
\frac{\langle \phi_{12}|H|\phi_{12}\rangle }{\langle \phi_{12}|\phi_{12}\rangle } = E_{o_1} + E_{o_2}+
\frac{\xi{(^{o_2}_{o_1}\left. \right.^{o_2}_{o_1}  )} - \xi^{in}{(^{o_2}_{o_1}\left. \right.^{o_2}_{o_1}  )} + \xi{(^{o_1}_{o_2}\left. \right.^{o_2}_{o_1}  )} - \xi^{in}{(^{o_1}_{o_2}\left. \right.^{o_2}_{o_1}  )}}{1-2\lambda_{o_1o_2;o_1o_2}},
\end{equation}
So that, if we use the fully symmetrized scattering defined in (\ref{symmxi}) and take a large sample volume in order for the $\lambda$ term to give a negligible contribution, this Hamiltonian mean value reduces, for $(o_1,o_2)\approx o$,  to
\begin{equation}
 \frac{\langle \phi_{12}|H|\phi_{12}\rangle }{\langle \phi_{12}|\phi_{12}\rangle }  = E_{o_1} + E_{o_2} + 2\frac{\hat{\xi}_{o_1o_2;o_1o_2}}{1 - \lambda_{_1o_2;o_1o_2}}
\approx 2E_o + 2\hat{\xi}_{oo;oo}.
\end{equation}

\subsection{Coherent superposition}

The third state of interest is the coherent superposition
$|\phi\rangle  = B^{\dagger 2}|0\rangle$, where
$B^\dagger  = a'B^\dagger _{o'} + a''B^\dagger _{o''}$, with $|a'|^2 +|a''|^2 = 1$ and $\vec{Q}_{o'} \neq \vec{Q}_{o''}$ in order for the Pauli and interaction scatterings to again cancel due to momentum conservation if the number of $o'$ cobosons in the ``in'' and ``out'' states are different.  As a direct consequence, $\langle \phi|\phi\rangle $ and $\langle \phi|H|\phi\rangle $ can only have terms in $|a'|^4,~|a''|^4$ and $|a'a''|^2$.  Using (\ref{III4}), the bare expansion of  $B^{\dagger 2}$ in terms of $B^{\dagger 2}_{o'},~B^{\dagger 2}_{o''}$ and $B^\dagger _{o'}B^\dagger _{o''}$, readily leads to the norm of the coherent state given by
\begin{equation}
%(III 10)
\langle \phi|\phi\rangle = |a'|^4L_{o'} + |a''|^4L_{o''} + |a'a''|^2L_{o',o''}
\end{equation}
with $L_{o'}= 2-2\lambda{(^{o'}_{o'}\left. \right.^{o'}_{o'}  )}$
and similarly for $L_{o''}$, while $
L_{o',o''}= 1-\lambda{(^{o''}_{o'}\left. \right.^{o''}_{o'}  )} - \lambda{(^{o'}_{o''}\left. \right.^{o''}_{o'}  )}$.
In terms of the symmetrized Pauli scatterings defined as in (\ref{symmxi}), this norm reduces to
\begin{equation}
%(III 11) 
\langle \phi|\phi\rangle  = 2 - 2|a'|^4\lambda_{o'o';o'o'} - 2|a''|^4\lambda_{o''o'';o''o''} - 8|a'a''|^2\lambda_{o'o'';o'o''}.
\end{equation}

In the same way, (\ref{cbH}) and (\ref{cbV}) allow us to write
\begin{equation}
%(III 12)  
\langle \phi|H|\phi\rangle  = |a'|^4A_{o'}  + |a''|^4A_{o''}+4|a'a''|^2A_{o',o''},
\end{equation}
where
$
A_{o'}=  2E_{o'}\left[2 - 2\lambda{(^{o'}_{o'}\left. \right.^{o'}_{o'}  )}\right] + 2\hat{\xi}{(^{o'}_{o'}\left. \right.^{o'}_{o'}  )} 
$
and similarly for $A_{o''}$, while
$$
A_{o',o''} =   (E_{o'} + E_{o''})\left[ 1 - \lambda{(^{o''}_{o'}\left. \right.^{o''}_{o'}  )} - \lambda{(^{o'}_{o''}\left. \right.^{o''}_{o'}  )}\right]  + \hat{\xi}{(^{o''}_{o'}\left. \right.^{o''}_{o'}  )} + \hat{\xi}{(^{o'}_{o''}\left. \right.^{o''}_{o'}  )} .
$$
For $o' \approx o'' \approx o$ in a large sample volume, the  expectation value of the Hamiltonian then reduces to
\begin{equation}
%(III 13)  
\frac{\langle \phi|H|\phi\rangle }{\langle \phi|\phi\rangle } \simeq 2E_o + 
 \frac{ 1 + 2|a'a''|^2)\hat{\xi}_{oo;oo}}{1 - (1 + 2|a'a''|^2)\lambda_{oo;oo}}
\approx 2E_o + \left(1 + 2|a'a''|^2\right)\hat{\xi}_{oo;oo}
%\eeq
\end{equation}

\subsection{Effective scattering for bosonized particles}
\label{sect.VD}

When compared to similar results for elementary bosons, namely equations (\ref{II3},\ref{II5},\ref{II8}), the Hamiltonian mean values in the three two-coboson states, obtained above, lead us to identify the diagonal effective scattering for elementary bosons $\bar{\xi}_{oo;oo}$ with the physically relevant combination of energylike scatterings defined in (\ref{I14}), namely
\begin{equation}
%(III 5'')  
 \bar{\xi}_{oo;oo}\equiv \hat{\xi}_{oo;oo} = \xi{(^{o}_{o}\left. \right.^{o}_{o}  )} - \xi^{in}{(^{o}_{o}\left. \right.^{o}_{o}  )}.
 \label{5-14}
\end{equation}
This effective sacttering contains a direct contribution as well as an exchange contribution which is symmetrical with respect to the ``in'' and ``out'' processes, as physically reasonable for $\xi^{in}{(^{o}_{o}\left. \right.^{o}_{o}  )} = \xi^{out}{(^{o}_{o}\left. \right.^{o}_{o}  )}$ due to equation (\ref{pscatt}).

For excitons or H atoms \cite{r15}, the diagonal direct scattering $\xi{(^{o}_{o}\left. \right.^{o}_{o}  )}$ reduces to 0, the repulsion between fermions $\alpha$ or fermions $\beta$ being as large as the attraction between $(\alpha,\beta)$.  On the opposite, the diagonal exchange scattering of these cobosons differ from zero: in 3D, it reads $\xi^{in}{(^{o}_{o}\left. \right.^{o}_{o}  )} = -({26\pi}/{3})R_0({a_B}/{L})^3$, where $R_0 = {\mu e^4}/{2\hbar^2\varepsilon^2} = e^2/2a_B$ is the coboson Rydberg energy.\cite{r15}

\section{Many elementary bosons}
\label{sect.many1}
\setcounter{equation}{0}

We now turn to states with a large number of bosons and first consider that these bosons are elementary bosons.

\subsection{Single state}

Let us start with the state having its $N$ bosons in a single state, $|\bar{\phi}_0\rangle  = \bar{B}^{\dagger N}_o|0\rangle $.  We physically expect the interaction term of the Hamiltonian expectation value to depend on the boson number through ${N(N - 1)}/{2}$ which corresponds to the number of interacting boson pairs $(o,o)$ we can form out of $N$ bosons in state $o$.  This leads us to expect that the Hamiltonian expectation value obtained for $N = 2$, as given in equation (\ref{II3}), must transform for general $N$ as
\beq
%(IV 1)  
\frac{\langle \bar{\phi}_0 \bar{H}|\bar{\phi}_0\rangle }{\langle \bar{\phi}_0|\bar{\phi}_0\rangle } = NE_o + \frac{N(N - 1)}{2}\bar{\xi}_{oo;oo}.
\label{IV1}
\eeq
For scatterings $\xi_{oo;oo}$ in $({a_B}/{L})^d$, this will induce a correction to the bare energy $NE_o$ of the order of $N\eta$, where $\eta$ is the dimensionless parameter associated to density given in (\ref{eta}).  Let us now recover this physically expected result.

Equation (\ref{mebB}) readily gives the well known normalization factor for elementary bosons, namely
\beq
%(IV 2) 
\langle \bar{\phi}_0|\bar{\phi}_0\rangle =\langle 0|\bar{B}^{N-1}_o\bar{B}_o\bar{B}^{\dagger N}_o|0\rangle =N\langle 0|\bar{B}^{N-1}_o\bar{B}^{\dagger  N-1}_o|0\rangle =N!
\eeq
Using the same (\ref{mebB}) on the one-body part of the Hamiltonian leads to
\beq
%(IV 3)  
\langle \bar{\phi}_0|\bar{H}_0|\bar{\phi}_0\rangle = \sum{E_i\left\{\langle \bar{\phi}_0|\bar{B}^\dagger _i\right\}\left\{\bar{B}_i|\bar{\phi}_0\rangle \right\}} = N^2E_o(N - 1)! = NE_o\langle \bar{\phi}_0|\bar{\phi}_0\rangle . 
\eeq
while on the two-body part, we find
\beq
%(IV 5)  
\langle \bar{\phi}_0|\bar{V}|\bar{\phi}_0\rangle  = \frac{1}{2}\sum{\bar{\xi}{(^{n}_{m}\left. \right.^{j}_{i}  )}\left\{\langle \bar{\phi}_0|\bar{B}^\dagger _m\bar{B}^\dagger _n\right\}\left\{\bar{B}_j\bar{B}_i|\bar{\phi}_0\rangle \right\}} = \frac{1}{2}N^2(N - 1)^2\bar{\xi}{(^{o}_{o}\left. \right.^{o}_{o}  )}(N-2)!
\eeq
This just proves that the guess of equation (\ref{IV1}) is correct.

\subsection{Fragmented state}

We now consider the fragmented state $|\bar{\phi}_{12}\rangle  = \bar{B}^{\dagger N_1}_{o_1}\bar{B}^{\dagger N_2}_{o_2}|0\rangle $. The same physical understanding leads us to expect the interaction term between elementary bosons $(o_1,o_2)$ of the Hamiltonian expectation value, as given in (\ref{II5}), to appear with a $N_1N_2$ prefactor.  Since we now have many bosons $o_1$ and many bosons $o_2$, we should also have an interaction term between bosons $o_1$ with a prefactor ${N_1(N_1 - 1)}/{2}$ and an interaction term between bosons $o_2$ with a prefactor ${N_2(N_2 - 1)}/{2}$.  From(\ref{II5}), we are thus led to guess
\begin{eqnarray}
%(IV 6) 
\frac{\langle \bar{\phi}_{12}|\bar{H}|\bar{\phi}_{12}\rangle }{\langle \bar{\phi}_{12}|\bar{\phi}_{12}\rangle } &=& N_1E_{o_1} + N_2E_{o_2} + \frac{N_1(N_1-1)}{2}\bar{\xi}_{o_1o_1;o_1o_1} + \frac{N_2(N_2-1)}{2}\bar{\xi}_{o_2o_2;o_2o_2} \nonumber\\
&&+2 N_1N_2\bar{\xi}_{o_1o_2;o_1o_2}.
\label{IV6}
\end{eqnarray}
For $o_1 \approx o_2 \approx o$, the Hamiltonian expectation value should thus be given by
\beq
%(IV 7)  
\frac{\langle \bar{\phi}_{12}|\bar{H}|\bar{\phi}_{12}\rangle }{\langle \bar{\phi}_{12}|\bar{\phi}_{12}\rangle } \simeq NE_o + \left(\frac{N(N-1)}{2} + N_1N_2\right)\bar{\xi}_{oo;oo}
\eeq
for $N_1 + N_2 = N$.  Let us now show this result explicitly.

Equation (\ref{mebB}) leads to\beq
%(IV 8)  
\bar{B}_i|\bar{\phi}_{12}\rangle  = \left(N_1\delta_{io_2}\bar{B}^{\dagger  N_1-1}_{o_1}\bar{B}^{\dagger  N_2}_{o_2} + N_2\delta_{io_2}\bar{B}^{\dagger  N_1}_{o_1}\bar{B}^{\dagger  N_2-1}_{o_2}\right)|0\rangle 
\label{IV8}
\eeq
so that, for $ o_1 \neq o_2$, we find
$ 
\bar{B}^{N_1}|\bar{\phi}_{12}\rangle  = N_1!\bar{B}^{\dagger  N_2}_{o_2}|0\rangle.
$
This readily gives
$ 
\langle \bar{\phi}_{12}|\bar{\phi}_{12}\rangle  = N_1!N_2!,
$
while
\beq
% (IV 11)  
\langle \bar{\phi}_{12}|\bar{H}_0|\bar{\phi}_{12}\rangle  = \sum{E_i{\langle \bar{\phi}_{12}|\bar{B}^\dagger _i}{\bar{B}_i|\bar{\phi}_{12}\rangle }} = E_{o_1}N_1^2(N_1-1)!N_2! + E_{o_2}N_2^2(N_2-1)!N_1!
\label{IV11}
\eeq
 If we now use equation (\ref{IV8}) to calculate $\bar{B}_j\bar{B}_i|\bar{\phi}_{12}\rangle $, the scalar product $\left\{\langle \bar{\phi}_{12}|\bar{B}^\dagger _m\bar{B}^\dagger _n\right\}$ $\left\{\bar{B}_j\bar{B}_i|\bar{\phi}_{12}\rangle \right\}$ which appears in $\langle \bar{\phi}_{12}|\bar{V}|\bar{\phi}_{12}\rangle $, leads to\beq
% (IV 12)  
\langle \bar{\phi}_{12}|\bar{V}|\bar{\phi}_{12}\rangle  = \frac{1}{2}(\bar{W}_{o_1}+\bar{W}_{o_2})+\bar{W}_{o_1,o_2}
\label{IV12}
\eeq
in which we have set
\beq
\bar{W}_{o_1} =
 \bar{\xi}{(^{o_1}_{o_1}\left. \right.^{o_1}_{o_1}  )}[N_1(N_1-1)]^2(N_1-2)!N_2!
\label{IV111}
\eeq
and similarly for $ \bar{W_{o_2}}$, while
\beq
\bar{W}_{o_1,o_2} = \biggl(\bar{\xi}{(^{o_2}_{o_1}\left. \right.^{o_2}_{o_1}  )} +\bar{\xi}{(^{o_1}_{o_2}\left. \right.^{o_2}_{o_1}  )}\biggr)(N_1N_2)^2(N_1-1)!(N_2-1)! 
\label{IV122}
\eeq
The expected result (\ref{IV6}) then readily follows from equations (\ref{IV11})-(\ref{IV12}).

\subsection{Coherent superposition}

 The third state of interest is the coherent superposition of states $|\bar{\phi}\rangle  = \bar{B}^{\dagger N}|0\rangle $ with $\bar{B}^\dagger  = a'\bar{B}^\dagger _{o'} +a'' \bar{B}^\dagger _{o''}$. Through similar physical arguments, we expect to have the interaction term in the energy for $N=2$ cobosons, as given in (\ref{II8}), to appear with a prefactor $N(N-1)/2$. To show it explicitly, we use equation (\ref{cohelmany}) to get
\beq
% (IV 13)  
\langle \bar{\phi}|\bar{\phi}\rangle  = \langle 0|\bar{B}^{N-1}(a'^*\bar{B}_{o'} + a''^*\bar{B}_{o''})\bar{B}^{\dagger N}|0\rangle  = N[|a'|^2 + |a''|^2]\langle 0|\bar{B}^{N-1}\bar{B}^{N-1}|0\rangle .
\eeq
 Its iteration, for $|a'|^2 + |a''|^2 = 1$, gives
$
\langle \bar{\phi}|\bar{\phi}\rangle  = N!
$
as for a condensate made of a single state.  In the same way, the norm of $\bar{B}_i|\bar{\phi}\rangle $ deduced from equation (\ref{cohelmany}), leads to
\beq
% (IV 15)  
\langle \bar{\phi}|\bar{H}_0|\bar{\phi}\rangle  = N(|a'|^2E_{o'} + |a''|^2E_{o''})N!,
\eeq
while the scalar product of the states $\langle \bar{\phi}|\bar{B}^\dagger _m\bar{B}^\dagger _n$ and $\bar{B}_i\bar{B}_j|\bar{\phi}\rangle $ leads, for $\vec{Q}_{o'} \neq \vec{Q}_{o''}$, to
\beq
% (IV 16)  
\langle \bar{\phi}|\bar{V}|\bar{\phi}\rangle  = \frac{1}{2}[N(N-1)]^2(N-2)!\left\{|a'|^4\bar{\xi}{(^{o'}_{o'}\left. \right.^{o'}_{o'}  )} + |a''|^2\bar{\xi}{(^{o''}_{o''}\left. \right.^{o''}_{o''}  )} + 2|a'a''|^2\left(\bar{\xi}{(^{o''}_{o'}\left. \right.^{o''}_{o'}  )} + \bar{\xi}{(^{o'}_{o''}\left. \right.^{o''}_{o'}  )}\right)\right\},
\eeq
which also reads
$$
 \langle \bar{\phi}|\bar{V}|\bar{\phi}\rangle  = \frac{N(N-1)}{2}[|a'|^4\bar{\xi}_{o'o';o'o'} + |a''|^2\bar{\xi}_{o''o'';o''o''} + 4|a'a''|^2\bar{\xi}_{o'o'';o'o''}]N! 
 $$
So that, for $o' \approx o'' \approx o'$, we end with the expected result, namely
\beq
\frac{\langle \bar{\phi}|H|\bar{\phi}\rangle }{\langle \bar{\phi}|\bar{\phi}\rangle } \simeq NE_o + \frac{N(N-1)}{2}\left(1 + 2|a'a''|^2\right)\bar{\xi}_{oo;oo}.
\eeq

 \section{Many composite bosons}
\label{sect.many2}
\setcounter{equation}{0}

 \subsection{Single state}

\begin{figure}
\begin{center}
\includegraphics[width=0.3\textwidth]{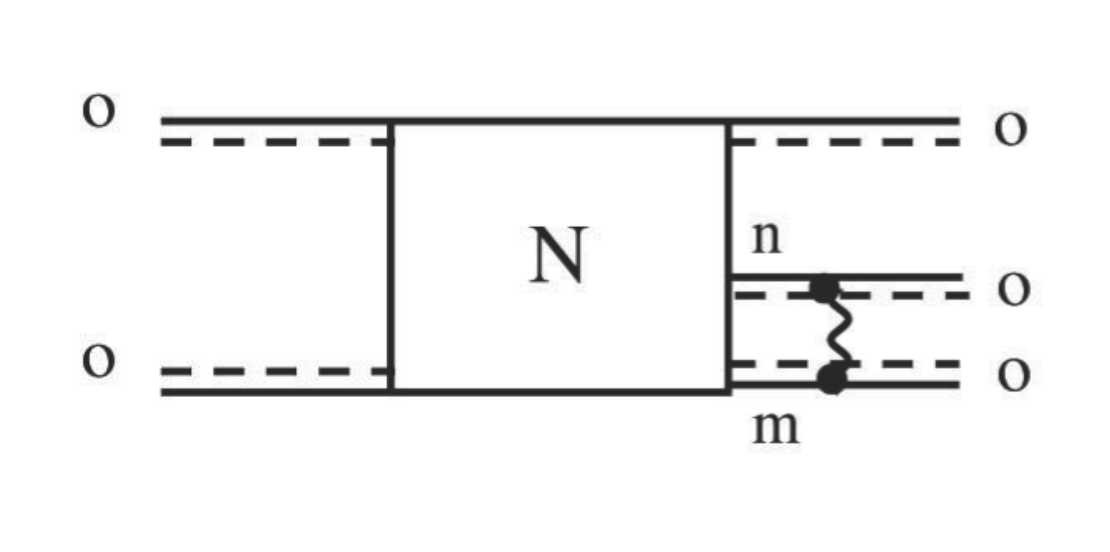}
\end{center}
\caption{Shiva diagram for a process having one fermion interaction between $N$ cobosons $o$, as given by the second term of equation (\ref{V2}).}
\label{fig6}
\end{figure}

The Hamiltonian mean value for the pure state $|\phi_0\rangle  = B^{\dagger N}_o|0\rangle $ has already been calculated in a previous work \cite{r15,r26}.  It has a naive contribution $NE_o$. It also has a set of density dependent corrections in $\eta^n$ with $n\geq1$, in contrast with elementary bosons which only have a $n=1$ term. This set of density terms comes from fermion exchanges between the N cobosons: Since the Hamiltonian expectation value $\langle H\rangle $ only contains one interaction by construction, the density terms for $n\geq2$ can only come from the fermion exchanges between 3 or more cobosons induced by the Pauli exclusion principle.  They are nicely visualized by Shiva diagrams with $n+1$ cobosons  and {\em one} interaction process between any two of these coboson lines (see Fig.~\ref{fig6}).
 Let us here repeat the main steps of this calculation for completeness - and also because the ones for $|\phi_{12}\rangle $ and $|\phi\rangle $ are conceptually similar, while far more complex.

\begin{figure}
\begin{center}
\includegraphics[width=0.75\textwidth]{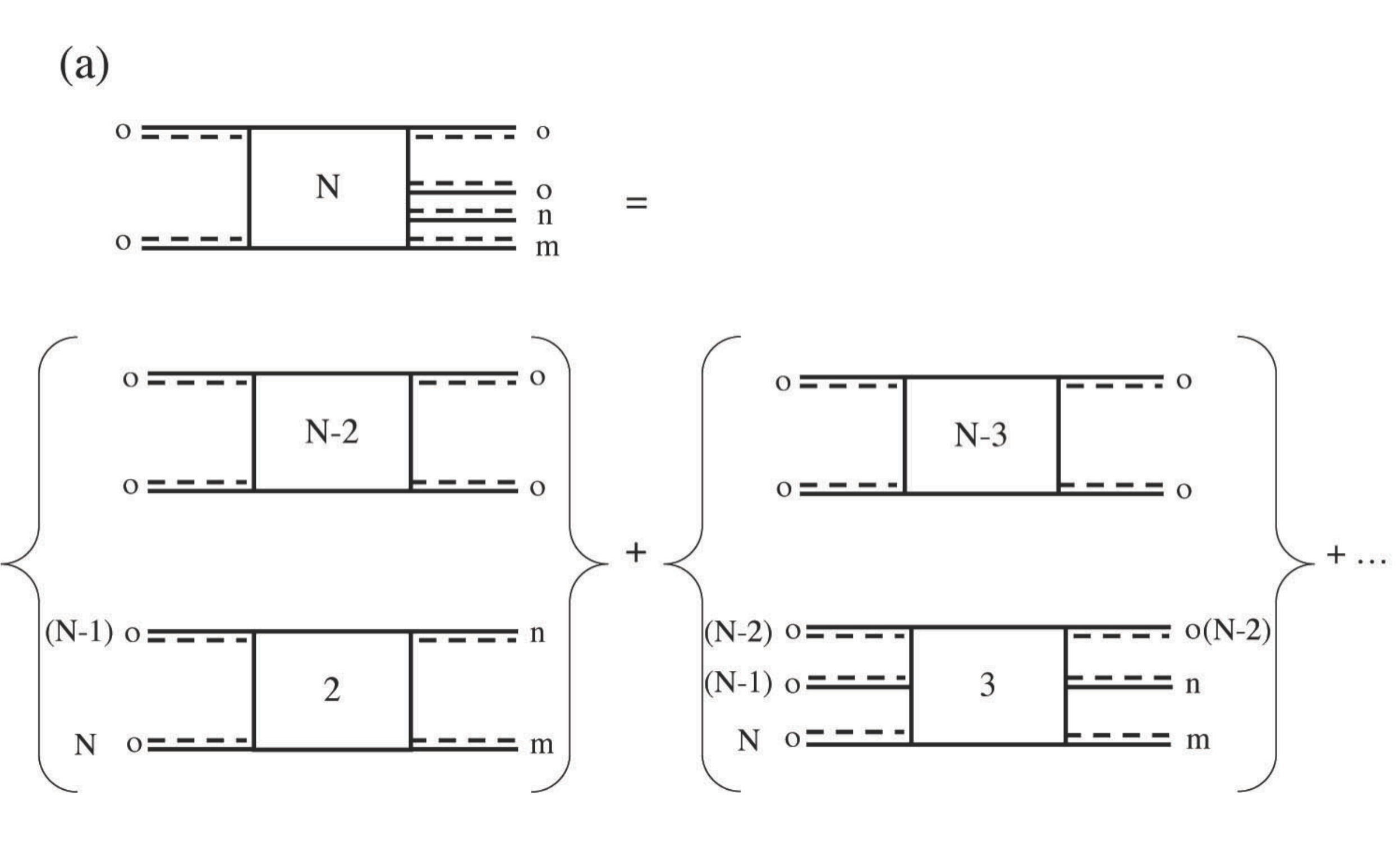}
\includegraphics[width=0.75\textwidth]{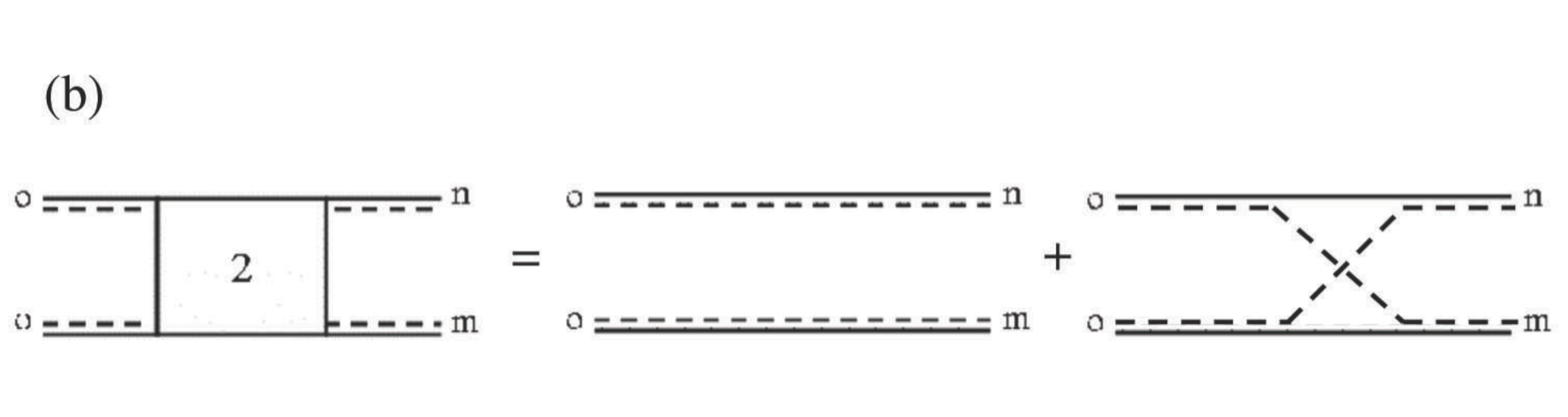}

\hspace{1.3cm}
\includegraphics[width=0.8\textwidth]{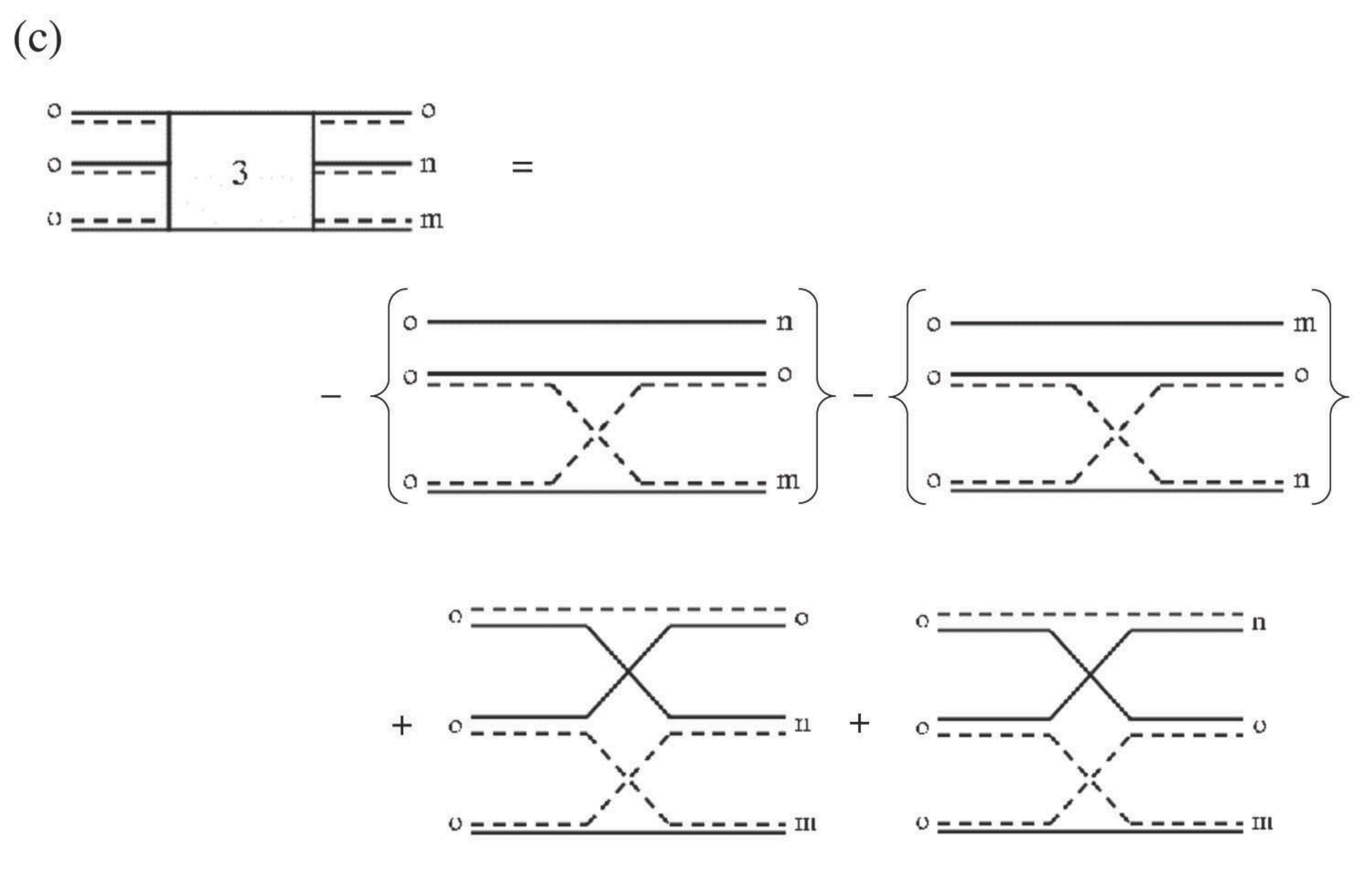}
\end{center}
\caption{a) Shiva diagram for the scalar product of $N$ coboson states appearing in equation (\ref{V3}), with $N$ cobosons $o$ on the left and $(N-2)$ cobosons $o$ plus two cobosons $(m,n)$ on the right. The standard way \protect\cite{r15,r17} to calculate this scalar product is to isolate $(N-2), (N-3),\ldots$ cobosons $o$ not involved in fermion exchanges with $(m,n)$.  The $N$ prefactors come from the number of ways to choose the coboson $o$ having fermion exchanges with $(m,n)$. These possible exchanges are shown in (b) and (c).}
\label{fig7}
\end{figure}
In order to calculate $\langle \phi_0|H|\phi_0\rangle $, we use one of the two key equations for coboson many body effects, namely (\ref{mcbH}), to find 
\beq
% (V 2)  
\langle \phi_0|H|\phi_0\rangle  = NE_o\langle \phi_0|\phi_0\rangle  + \frac{N(N-1)}{2}\sum{\xi{(^{n}_{m}\left. \right.^{o}_{o}  )}}\langle 0|B^N_oB^{\dagger  N-2}_oB^\dagger _mB^\dagger _n|0\rangle . 
\label{V2}
\eeq
The second term of this equation is shown in Fig.~\ref{fig6}. In it, appears the scalar product of N coboson states with two cobosons different from $o$ on the right (see Fig.~\ref{fig7}).  This scalar product is calculated using another key equation for many-body effects, namely (\ref{mcbB}). This equation leads to
\beq
\label{V3} 
\langle 0|B^{N-2}_oB_mB_nB^{\dagger N}_o|0\rangle  = \langle 0|B^{N-2}_oB_m\left[ N\delta_{no}B^{\dagger  N-1}_0 - N(N-1)B^{\dagger  N-2}_o\sum{\lambda{(^{p}_{n}\left. \right.^{o}_{o}  )}B^\dagger _p}\right]|0\rangle .
\eeq
We then use the same commutation (\ref{mcbB}) to pass $B_m$ over $B^{\dagger  N-1}_o$ and $B^{\dagger  N-2}_o$.  Since $\left[\lambda{(^{n}_{m}\left. \right.^{i}_{j}  )}\right]^* = \lambda{(^{j}_{i}\left. \right.^{n}_{m}  )}$, this allows us to expand the above scalar product as (see Fig.~\ref{fig7})
\begin{eqnarray}
% (V 4)  
&&\langle 0|B^N_oB^\dagger _mB^\dagger _nB^{\dagger  N-2}_o|0\rangle  = \nonumber \\
&&\hspace{1cm} N(N-1)\left[\delta_{on}\delta_{om} - \lambda{(^{o}_{o}\left. \right.^{n}_{m}  )}\right](N-2)!F_{N-2}  \nonumber\\
&&\hspace{1cm}+N(N-1)(N-2)^2 
\left[-\delta_{on}\lambda{(^{o}_{o}\left. \right.^{o}_{m}  )}-\delta_{om}\lambda{(^{o}_{o}\left. \right.^{o}_{m}  )} + \lambda_3 + \lambda_3'\right]
(N-3)! F_{N-2} + \ldots.\nonumber\\
\label{V4}
\end{eqnarray}
where $\lambda_3$ and $\lambda_3'$ are three-leg scatterings shown in Fig.~\ref{fig7}(c).
$$
\lambda_3 = \lambda \left( \begin{array}{cc}
o & o \\
o & n \\
o & m \end{array} \right), \hspace{1cm} 
\lambda_3' =  \lambda \left( \begin{array}{cc}
o & n \\
o & o \\
o & m \end{array} \right).
$$
in which the cobosons $(m,n)$ exchange their fermions with one coboson $o$ to produce three cobosons $o$. This expansion actually follows the standard procedure \cite{r17} to calculate scalar products, namely, we first isolate terms in
$$
 \langle 0|B^{N-P}_oB^{\dagger  N-P}_o|0\rangle  = (N-P)!F_{N-P}, 
 $$ 
 with $P \geq 2$. We then connect the remaining cobosons in all possible ways while enforcing the cobosons $o$ to be ``never alone''  as in Fig.~\ref{fig7}(b),(c).  The $N$ prefactors in Fig.~\ref{fig7}(a) correspond to the number of ways we can choose the cobosons $o$ among $N$ on the left and among $(N-2)$ on the right.  In the case of the first term of this figure, we just have to choose the two cobosons $o$ on the left; this is why this first term appears with a prefactor $N(N-1)$. In the second term, we also have to choose a coboson $o$ on the right and a third coboson $o$ on the left. This is why this term appears with a prefactor $[N(N-1)(N-2)] (N-2)$. And so on ...

By inserting equation (\ref{V4}) into (\ref{V2}), we readily find
\beq
% (V 5) 
\frac{\langle \phi_0|H|\phi_0\rangle }{\langle \phi_0|\phi_0\rangle } = NE_o + \frac{F_{N-2}}{F_N} \frac{N(N-1)}{2} \left[\xi{(^{o}_{o}\left. \right.^{o}_{o}  )} - \xi^{in}{(^{o}_{o}\left. \right.^{o}_{o}  )}\right] + \ldots
\eeq
so that, since ${F_{N-2}}/{F_N} = ({F_{N-2}}/{F_{N-1}})({F_{N-1}}/{F_N}) = 1 + {\cal O}(\eta)$,
we end for large $N$ with
\beq
% (V 6)  
\frac{\langle \phi_0|H|\phi_0\rangle }{\langle \phi_0|\phi_0\rangle } = N\left[E_o + \frac{N}{2}\hat{\xi}_{oo;oo} +{\cal  O}(\eta^2 )\right],
\eeq
This shows that the first correction to the bare energy $E_o$ is of order $\eta = N(a_B/L)^d$, because, as shown in Section V.D, $\hat{\xi}_{oo;oo}$ is of the order of $({a_B}/{L})^d$. Note that, as for Coulomb interaction $\xi{(^{o}_{o}\left. \right.^{o}_{o}  )} =0$, the effective scattering $\hat{\xi}_{oo;oo}$ reduces to $-\xi^{in}{(^{o}_{o}\left. \right.^{o}_{o}  )}$ which for 3D excitons  is equal to $ -({26\pi}/{3})R_0({a_B}/{L})^3$. Consequently, this Hamiltonian mean value is just the energy of $N$ electron-hole pairs obtained by Keldysh and Kozlov \cite{t1}, using a completly different approach in which these pairs are not treated as coboson entities, as we do here.  

 \subsection{Coherent superposition}

In order to calculate the mean value of the Hamiltonian for  the N coherent cobosons $| \phi\rangle =B^{\dagger N}| 0 \rangle$, we use the results derived in Section \ref{sect.cohbos}. From (\ref{V18}),  this mean value appears as
\beq
% (V 23)  
\langle 0|B^NHB^{\dagger N}|0\rangle  = NE_{o}\langle 0|B^N\tilde{B}^\dagger B^{\dagger  N-1}|0\rangle  + \frac{N(N-1)}{2}\Delta
\label{V23}
\eeq
\beq
\Delta = \sum{\xi_{mn}\langle 0|B^NB^\dagger _mB^\dagger _nB^{\dagger  N-2}|0\rangle }.
\eeq
where ${\xi_{mn}}$ is given in eq(3-28). The first term of (\ref{V23}) reduces to $NE_{o}\langle \phi|\phi\rangle $ for $E_{o'} \approx E_{o''} \approx E_{o}$ since $\tilde{B}^\dagger  \approx B^\dagger$.  
To calculate $\Delta$, we use equation (\ref{gmcbD}) for $\langle 0|B^NB^\dagger _m$.  This leads to
\beq
% (V 24)  
\Delta = N\sum{\xi_{mn}\delta^*_m\langle 0|B^{N-1}B^\dagger _nB^{\dagger  N-2}|0\rangle }- N(N-1)\sum{\xi_{mn}\lambda^*_{mp}\langle 0|B^{N-2}B_pB^\dagger _nB^{\dagger  N-2}|0\rangle }
\eeq
 In the first term, we again use (\ref{Bmult}) for $\langle 0|B^{N-1}B^\dagger _n$, while we use the commutator $[B_p,B^\dagger _n]$ given in (\ref{cbB'}) to calculate the second term of $\Delta$.  This leads to
\beq
% (V 25)  
\Delta = N(N-1)(N-2)!G_{N-2}\sum{\xi_{mn}(\delta_m^*\delta_n^* - \lambda_{mn}^*)} + ... G_{N-3} + \ldots
\label{V25}
\eeq
If we now use the definitions of $\delta_m,~\xi_{mn}$ and $\lambda_{mn}$ given in equations (\ref{2-7}), (\ref{gcblam}), and (\ref{V16}), we find that for $\vec{Q}_{o'} \neq \vec{Q}_{o''}$, the sum in (\ref{V25}) reduces to
$$
% (V 26)
|a'|^4\hat{\xi}(^{o'}_{o'}\left. \right.^{o'}_{o'}  ) + |a''|^4\hat{\xi}{(^{o''}_{o''}\left. \right.^{o''}_{o''}  )}
+ 2|a'a''|^2\left\{\hat{\xi}{(^{o''}_{o'}\left. \right.^{o''}_{o'}  )} + \hat{\xi}{(^{o''}_{o'}\left. \right.^{o'}_{o''}  )}\right\}$$
with $\hat{\xi} = \xi - \xi^{in}$ as defined in equation (\ref{I14}).  By collecting all the terms, we end with
\begin{eqnarray}
% (V 27)  
\frac{\langle \phi|H|\phi\rangle }{\langle \phi|\phi\rangle }&\simeq& 
NE_{o}  + \frac{N(N-1)}{2}\frac{G_{N-2}}{G_N}
\left\{|a'|^4\hat{\xi}_{o'o';o'o'} + |a''|^4\hat{\xi}_{o''o'';o''o''} + 4|a'a''|^2\hat{\xi}_{o'o'';o'o''}\right\} + \ldots \nonumber\\
\end{eqnarray}
within corrections of the order of $(E_{o'} - E_{o''})$. For $o' \approx o'' \approx \ o$ and $N$ large, the above result reduces to
\beq
% (V 28)  
\frac{\langle \phi|H|\phi\rangle }{\langle \phi|\phi\rangle }=  N[E_o 
+ \frac{N}{2}(1 + 2|a'a''|^2)\hat{\xi}_{oo;oo} 
+ O|\eta^2)].
\eeq

 \subsection{Fragmented state}
 The last N-coboson state we must consider $|\phi_{12}\rangle  = B^{\dagger N_1}_{o_1}B^{\dagger N_2}_{o_2}|0\rangle $ has two large numbers of different cobosons in state $o_1$ and $o_2$. The very many exchanges which exist within the $o_1$ population, within the $o_2$ population, and between the $o_1$ and $o_2$ populations make this many-body calculation quite tricky. This is why we have kept it for the end, the previous ones having the role of useful exercises. In view of the above results, we can however guess that the Hamiltonian mean value for composite bosons in this fragmented state should read as the one for elementary bosons, namely (\ref{IV6}), with, according to (\ref{5-14}), $\bar{\xi}_{o_1o_1;o_1o_1}$ and $\bar{\xi}_{o_1o_2;o_1o_2}$ replaced by $\hat{\xi}_{o_1o_1;o_1o_1}$ and $\hat{\xi}_{o_1o_2;o_1o_2}$, respectively.  Let us show this nicely simple explicitly.
 
\begin{figure}
\begin{center}
\includegraphics[width=0.7\textwidth]{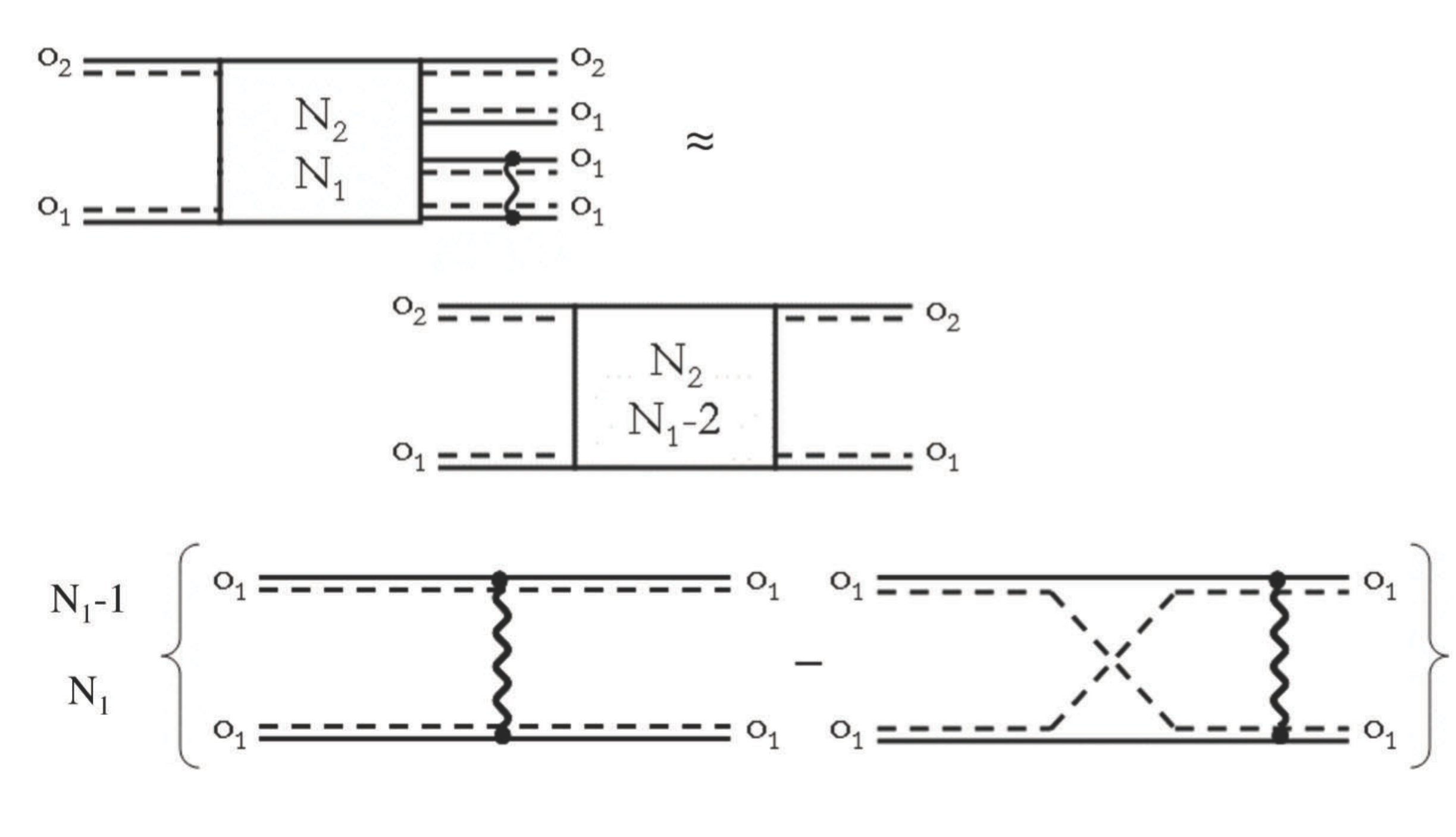}
\end{center}
\caption{Shiva diagrams for $\Delta_{11}$ defined in equation (\ref{V31}), in which two cobosons $o_1$ among $N_1$ have an interaction scattering, the other $N_2$ cobosons $o_2$ just possibly exchanging their fermions with the cobosons $o_1$. The $N$ prefactors are the number of ways to choose the cobosons $o_1$ on the left.}
\label{fig8}
\end{figure}

\begin{figure}
\begin{center}
\includegraphics[width=0.8\textwidth]{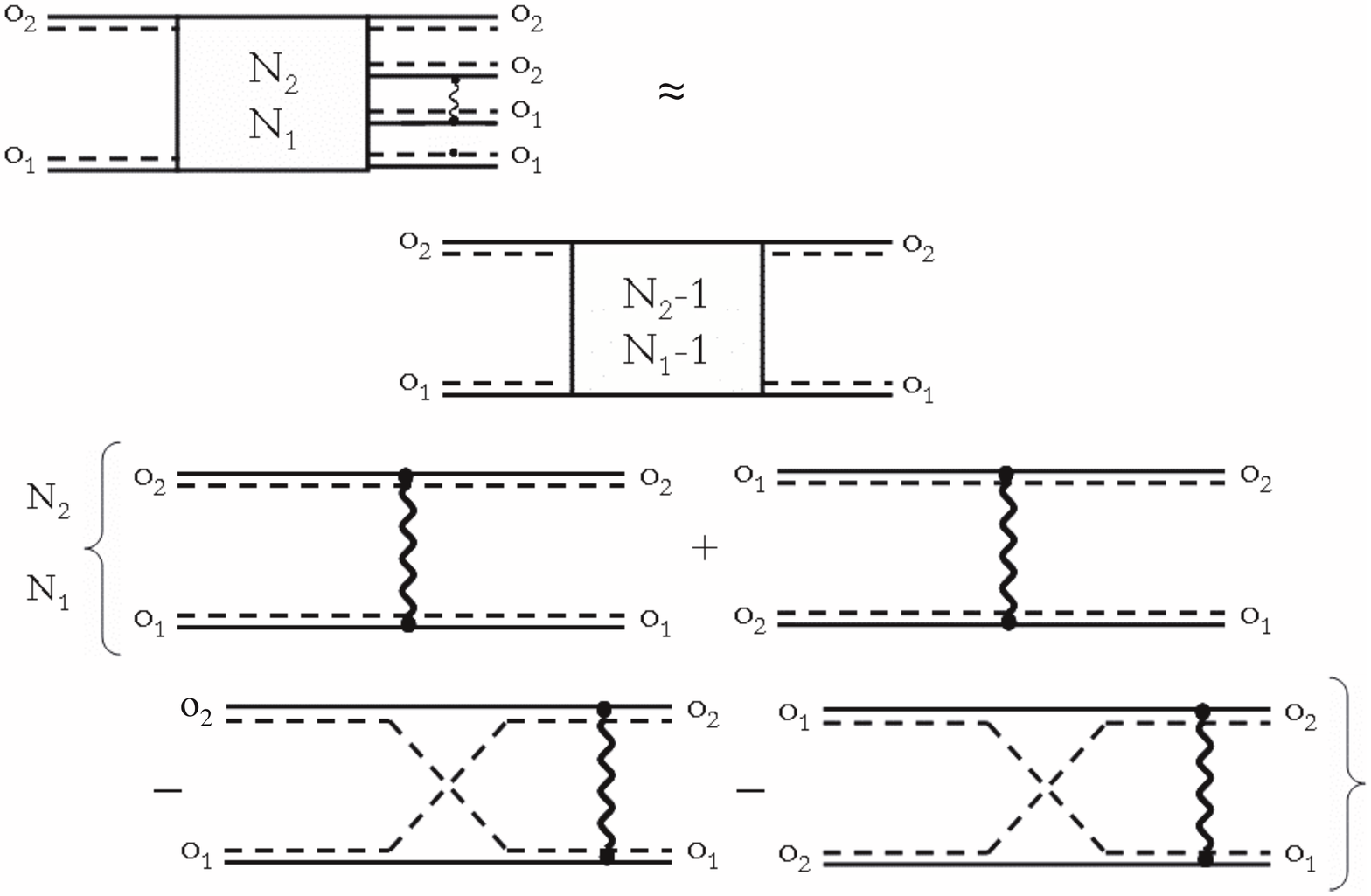}
\end{center}
\caption{Same as Fig.~8, for $\Delta_{12}$ defined in equation (\ref{V32}), the interaction taking place between one coboson $o_1$ among $N_1$ and one coboson $o_2$ among $N_2$. The $N$ prefactors are the number of ways to choose the cobosons $(o_1,o_2)$ on the left.}
\label{fig9}
\end{figure}

 We first use (\ref{mcbH}) in the coboson many-body effect section, to replace $HB^{\dagger N_1}_{o_1}$ in $H|\phi_{12}\rangle$. This leads to 
\beq
% (V 29)  
H|\phi_{12}\rangle  = \left\{B^{N_1}_{o_1}H + N_1E_{o_1}B^{\dagger N_1}_{o_1} + N_1B^{\dagger  {N_1}-1}_{o_1}V^\dagger _{o_1} + \frac{N_1(N_1-1)}{2}B^{\dagger  N_1-2}_{o_1} \sum{\xi{(^{n}_{m}\left. \right.^{o_1}_{o_1}  ) }B^\dagger _mB^\dagger _n}\right\}B^{\dagger N_2}_{o_2}|0\rangle .
\eeq
We again use (\ref{mcbH}) for $HB^{\dagger N_2}_{o_2}|0\rangle$ and (\ref{mcbV}) for $V^\dagger _{o_1}B^{\dagger N_2}_{o_2}|0\rangle $. This allows to split $\langle \phi_{12}|H|\phi_{12}\rangle $ into
\beq
 %(V 30) 
 \langle \phi_{12}|H|\phi_{12}\rangle  = (N_1E_{o_1} + N_2E_{o_2}) \langle \phi_{12}|\phi_{12}\rangle  + \frac{N_1(N_1-1)}{2}\Delta_{11} + \frac{N_2(N_2-1)}{2}\Delta_{22} + N_1N_2\Delta_{12}.
 \eeq
$\Delta_{11}$, shown in Fig.~\ref{fig8}, describes the interactions of two among $N_1$ cobosons in state $o_1$, the other $N_2$ cobosons $o_2$ possibly having fermion exchanges with the cobosons $o_1$. The precise value of $\Delta_{11}$ is
\beq
% (V 31) 
\Delta_{11} = \sum{\langle 0|B^{N_2}_{o_2}B^{N_1}_{o_1}B^\dagger _mB^\dagger _nB^{\dagger  N_1-2}_{o_1}B^{\dagger  N_2}_{o_2}|0\rangle \xi{(^{n}_{m}\left. \right.^{o_1}_{o_1}  ) }},
\label{V31}
\eeq
and similarly for $\Delta_{22}$, while $\Delta_{12}$, shown in Fig.~\ref{fig9}, results from the interaction of one coboson $o_1$ with one coboson $o_2$. It reads
\beq
% (V 32) 
\Delta_{12} = \sum{\langle 0|B^{N_2}_{o_2}B^{N_1}_{o_1}B^\dagger _mB^\dagger _nB^{\dagger  N_1-1}_{o_1}B^{\dagger  N_2-1}_{o_2}|0\rangle \xi{(^{n}_{m}\left. \right.^{o_2}_{o_1}  ) }}.
\label{V32}
\eeq

The reader who is knowledgable about Shiva diagrams \cite{r15,r17} will immediately see from Fig.~8 that, for $\vec{Q}_{o_1} \neq \vec{Q}_{o_2}$, i.e. for $\xi{(^{o_2}_{o_1}\left. \right.^{o_1}_{o_1}  ) } = 0 = \xi^{in}{(^{o_2}_{o_1}\left. \right.^{o_1}_{o_1}  ) }$, we must have
\beq
% (V 33) 
\Delta_{11} \simeq N_1(N_1-1)\left[\xi{(^{o_1}_{o_1}\left. \right.^{o_1}_{o_1}  ) } - \xi^{in}{(^{o_1}_{o_1}\left. \right.^{o_1}_{o_1}  )}\right](N_1-2)!N_2!G_{N_1-2,N_2},
\eeq
and similarly for $\Delta_{22}$. Here $G_{N_1,N_2}$ is defined as $F_N$, through the norm of the $B^{\dagger N_1}_{o_1}B^{\dagger N_2}_{o_2}|0\rangle $ state (see equation (\ref{g12norm})).

In the same way, the Shiva diagram of Fig.~\ref{fig9} readily leads to, for $\vec{Q}_{o_1} \neq \vec{Q}_{o_2}$,
\beq
% (V 35)  
\Delta_{12} \simeq N_1N_2\left[[\xi{(^{o_2}_{o_1}\left. \right.^{o_2}_{o_1}  )} + \xi{(^{o_1}_{o_2}\left. \right.^{o_2}_{o_1}  )} -\xi^{in}{(^{o_2}_{o_1}\left. \right.^{o_2}_{o_1}  )} - \xi^{in}{(^{o_1}_{o_2}\left. \right.^{o_2}_{o_1}  )}\right]
(N_1-1)!(N_2-1)!G_{N_1-1,N_2-1}.
\eeq
By collecting all these terms, we end with
\begin{eqnarray}
% (V 36)  
\frac{\langle {\phi}_{12}|{H}|{\phi}_{12}\rangle }{\langle {\phi}_{12}|{\phi}_{12}\rangle } &\simeq & N_1E_{o_1} + N_2E_{o_2} + \frac{N_1(N_1-1)}{2}\hat{\xi}_{o_1o_1;o_1o_1}\frac{G_{N_1-2,N_2}}{G_{N_1,N_2}} + \frac{N_2(N_2-1)}{2}\hat{\xi}_{o_2o_2;o_2o_2}\frac{G_{N_1,N_2-2}}{G_{N_1,N_2}} \nonumber\\
&&+ 2N_1N_2\hat{\xi}_{o_1o_2;o_1o_2}\frac{G_{N_1-1,N_2-1}}{G_{N_1,N_2}} \label{V36}
\end{eqnarray}
Like  $F_N$, the factor $G_{N_1,N_2}$, which comes from the many fermion exchanges which take place between the $N_1$ cobosons  in state $o_1$ and the $N_2$ cobosons in state $o_2$, is exponentially small.  However the ratios of $G_{N_1,N_2}$'s are nearly 1 at lowest order in density.
Enforcing $N_1 + N_2 = N$ and $o_1 \approx o_2 \approx o$, we ultimatly find the expected result, namely
\beq
% (V 37) 
\frac{\langle {\phi}_{12}|{H}|{\phi}_{12}\rangle }{\langle {\phi}_{12}|{\phi}_{12}\rangle } \simeq NE_o +\left(\frac{N(N-1)}{2} + N_1N_2\right)\hat{\xi}_{oo;oo}.
\eeq

Let us end this section by returning to the interaction parts $\Delta_{11},~\Delta_{22}$ and $\Delta_{12}$ of the Hamiltonian mean value defined in equations (\ref{V31}) and (\ref{V32}) and by calculating them, not through Shiva diagrams, but in a pedestrian way using the commutators appropriate to many body effects, namely equations (\ref{mcbB}) and (\ref{mcbD}).  The problem is to get the scalar products of coboson states which appear in these $\Delta$'s. This always is the tricky part of any calculation involving cobosons.  The ones of interest here are
\beq
% (V 45)  
S^{(P)} = \langle 0|B^{N_2}_{o_2}B^{N_1}_{o_1}B^\dagger _mB^\dagger _nB^{\dagger ~N_1-1-P}_{o_1}B^{\dagger ~N_2-1+P}_{o_2}|0\rangle 
\eeq
 for $P = (0,\pm1)$.  This calculation is done along a line similar to the one we have used when we only had one type of cobosons, namely we isolate the norm of states like $B^{\dagger N'_1}_{o_1}B^{\dagger N'_2}_{o_2}|0\rangle $ with $(N'_1,N'_2)$ decreasing from $(N_1-1-P,~N_2-1+ P)$.  To do it, we first use equation (\ref{mcbB}) to rewrite $B^{N_1}_{o_1}B^\dagger _m$.  This leads to four terms.  We then use (\ref{mcbB}) to get $\langle 0|B^{N_2}_{o_2}B^\dagger _m$ and equation (\ref{mcbD}) to get $\langle 0|B^{N_2}_{o_2}D_{om}$.  This allows to split $S^{(P)}$ as 
\beq
% (V 46) 
S^{(P)} = N_1T^{(P)}_1 + N_2T^{(P)}_2 - N_1(N_1-1)U^{(P)}_{11} - N_2(N_2-1)U^{(P)}_{22} - N_1N_2U^{(P)}_{12},
\eeq
where the contributions $T^{(P)}_1$ and $T^{(P)}_2$ are somewhat direct since they read
\begin{eqnarray}
% (V 47) 
T^{(P)}_1 &=& \delta_{o_1m}\langle 0|B^{N_2}_{o_2}B^{N_1-1}_{o_1}|\psi_{n,P}\rangle \label{V47} \\
% (V 48)  
T^{(P)}_2 &=& \delta_{o_2m}\langle 0|B^{N_1}_{o_1}B^{N_2-1}_{o_2}|\psi_{n,P}\rangle .
\end{eqnarray}
where we gave set $|\psi_{n,P}\rangle =B^\dagger _nB^{\dagger ~N_1-1-P}_{o_1}B^{\dagger ~N_2-1+P}_{o_2}|0\rangle $

The three other terms $U^{(P)}_{11},~U^{(P)}_{22}$ and $U^{(P)}_{12}$ contain one Pauli scattering explicitly which describes the fermion exchanges between two cobosons in state $o_1$, two cobosons in state $o_2$, and one coboson in state $o_1$ with one coboson in state $o_2$, as understood from their $N$ prefactors.  These $U^{(P)}$ terms precisely are
\begin{eqnarray}
% (V 49) 
U^{(P)}_{11} &=& \sum\limits_{i}{\lambda{(^{o_1}_{o_1}\left. \right.^{i}_{m}  )}\langle 0|B^{N_1-2}_{o_1}B^{N_2}_{o_2}B_i|\psi_{n,P}\rangle }\\
U^{(P)}_{22} &=& \sum\limits_{i}{\lambda{(^{o_2}_{o_2}\left. \right.^{i}_{m}  )}\langle 0|B^{N_2-2}_{o_2}B^{N_1}_{o_1}B_i|\psi_{n,P}\rangle }\\
U^{(P)}_{12} &=& \sum\limits_{i}{\left[\lambda{(^{o_2}_{o_1}\left. \right.^{i}_{m}  )} + \lambda{(^{o_1}_{o_2}\left. \right.^{i}_{m}  )}\right]}
\langle 0|B^{N_2-1}_{o_2}B^{N_1-1}_{o_1}B_i|\psi_{n,P}\rangle
\end{eqnarray}
Their leading contributions in fermion exchanges are obtained by passing $B_i$ over $B^\dagger _n$ in $|\psi_{n,P}\rangle$ through the commutator (\ref{cbB'}).  The trivial term corresponds to taking $i = n$ while the two other terms generate additional exchanges between $i$, or $n$, and the other cobosons in states  $o_1$ and $o_2$.  In the case of $U^{(P)}_{11}$, the remaining matrix element is just $(N_1-2)!N_2!G_{N_1-2,N_2}$ for $P = 1$ while it contains additional Pauli scatterings for $P \neq 1$.  In the same way, the remaining matrix element of $U^{(P)}_{12}$ is just $(N_1-1)!(N_2-1)!G_{N_1-1,N_2-1}$ for $P = 0$, while for other P's, this matrix element contains additional Pauli scatterings.  Consequently, the contributions to the $U^{(P)}$'s with only one Pauli scattering reduce to
\begin{eqnarray}
% (V 50)  
U^{(1)}_{11} &\simeq& \lambda(^{o_1}_{o_1}\left. \right.^{n}_{m}  )(N_1-2)!N_2!G_{N_1-2,N_2} \\
U^{(0)}_{12} &\simeq& \left[\lambda{(^{o_2}_{o_1}\left. \right.^{n}_{m}  )} + \lambda{(^{o_1}_{o_2}\left. \right.^{n}_{m}  )}\right](N_1-1)!(N_2-1)!G_{N_1-1,N_2-1} \nonumber
\end{eqnarray}
and similarly, $U^{(1)}_{22}$ obtained from $U^{(1)}_{11}$ by changing 1 into 2.

Let us now turn to the $T^{(P)}_1$ term defined in (\ref{V47}).  For $P = 1$, we pass $B^\dagger _n$ over $B^{N_1-1}_{o_1}$ using the many-body commutator (\ref{mcbB}). The contribution without exchange, which is the dominant one at small density, reads
\beq
% (V 51) 
T^{(1)}_1 \simeq (N_1-1)\delta_{o_1m}\delta_{o_1n}(N_1-2)!N_2!G_{N_1-2~N_2}.
\eeq
If for $P = 0$, we do the same but with $B^{N_2}_{o_2}$, we find
\beq
% (V 52) 
T^{(0)}_1 \simeq N_2\delta_{o_1m}\delta_{o_2n}(N_1-1)!(N_2-1)!G_{N_1-1~N_2-1}.
\eeq
 On the opposite, for $P = -1$, additional exchange processes are necessary to transform the $B^{\dagger N_1}_{o_1}$ operator on the right into $B^{\dagger ~N_1-1}_{o_1}$ 
in order to get rid of the $B^{N_1-1}_{o_1}$ operator of the left, as necessary to generate a $G_{N_1-1,N_2}$ factor.
By calculating $T^{(P)}_2$ along the same line and by inserting all these matrix elements into $\Delta_{11}$ and $\Delta_{12}$ given in equations (\ref{V31},\ref{V32}), we end with the Hamiltonian mean value written in equation (\ref{V36}).

\section{Discussion}
\label{sect.concl}
\setcounter{equation}{0}

In the previous sections, we have performed detailed calculations of the Hamiltonian mean value  for three different types of states, namely a pure state $|\phi_0\rangle  = B^{\dagger N}_o|0\rangle $,  a coherent state $|\phi\rangle  = (a'B^\dagger _{o'} + a''B^\dagger _{o''})^N|0\rangle $ and a fragmented state $|\phi_{12}\rangle  = B^{\dagger N_1}_{o_1}B^{\dagger N_2}_{o_2}|0\rangle $ with $N_1 + N_2 = N$.  We have taken $N = 2$ first, since the calculations are rather trivial and most of the physics can already be understood from this two-body problem.  As explicitly shown, the results for large $N$ can be deduced from the ones for $N = 2$ by putting N in front of the scatterings, in this way producing a density correction in $\eta = N({a_B}/{L})^d$ to the bare energy, as physically reasonable.

We have considered elementary bosons as well as composite bosons.  The calculations with composite bosons are done using the new many-body theory designed for them, in which the fermion exchanges appear explicitly through a set of Pauli scatterings $\lambda{(^{n}_{m}\left. \right.^{j}_{i}  )}$.  These exchanges are visualized by Shiva diagrams, which are of great help to drive the algebra in the right direction in case of many body effects - the last section being quite convincing with respect to the utility of these diagrams.  The Hamiltonian used when the composite nature of the particles is retained, is the microscopic Hamiltonian for fermions.  On the opposite, an effective Hamiltonian is necessary when the composite bosons are replaced by elementary particles.  In order to recover the exact composite boson results from this effective Hamiltonian, for the simple problem in which only one interaction scattering appears---the Hamiltonian mean value being first order in the interaction by construction---we must adjust the diagonal scattering of this effective Hamiltonian to be such that
\beq
% (VI 1) 
\bar{\xi}_{ij;ij} \equiv \hat{\xi}_{ij;ij}
\eeq
where $\hat{\xi}_{ij;ij}$ is the symmetrical combination of direct and exchange interaction scatterings of two composite bosons, defined as
\beq
% (VI 2)  
\hat{\xi}_{ij;ij} = \frac{1}{2}\left[\xi{(^{j}_{i}\left. \right.^{j}_{i}  )} + \xi{(^{i}_{j}\left. \right.^{j}_{i}  )}\right] - \frac{1}{2}\left[\xi^{in}{(^{j}_{i}\left. \right.^{j}_{i}  )} + \xi^{in}{(^{i}_{j}\left. \right.^{j}_{i}  )}\right].
\eeq
The term $\xi{(^{n}_{m}\left. \right.^{j}_{i}  )}$ corresponds to the interaction scattering of the composite boson many-body theory which describes fermion interactions without fermion exchange, while the term $\xi^{in}{(^{n}_{m}\left. \right.^{j}_{i}  )} = \sum{\lambda{(^{n}_{m}\left. \right.^{q}_{p}  )}\xi(^{q}_{p}\left. \right.^{j}_{i}  )}$ 
corresponds to the exchange interaction scattering, with $\lambda{(^{n}_{m}\left. \right.^{q}_{p}  )}$ describing fermion exchange in the absence of fermion interaction.  Note that, in the case of diagonal scattering, we do have $\xi^{in}{(^{j}_{i}\left. \right.^{j}_{i}  )} = \xi^{out}{(^{j}_{i}\left. \right.^{j}_{i}  )}$, as physically required by the time reversal of the scattering.

Within this identification of the diagonal effective scattering for bosonized particles, we find that the Hamiltonian mean values read the same for elementary and composite bosons in the three states of interest, namely
\begin{eqnarray}
 \frac{\langle \phi_0|H|\phi_0\rangle }{\langle \phi_0|\phi_0\rangle } & \simeq& NE_0 + \frac{N(N-1)}{2}\hat{\xi}_{oooo} \label{8-3}\\
\frac{\langle \phi|H|\phi\rangle }{\langle \phi|\phi\rangle } & \simeq&  NE_0 + \frac{N(N-1)}{2}  (1 + 2|a'a''|^2)\hat{\xi}_{oooo} \label{8-4}\\
\frac{\langle \phi_{12}|H|\phi_{12}\rangle }{\langle \phi_{12}|\phi_{12}\rangle } & \simeq& NE_0 + \left(\frac{N(N-1)}{2} + N_1N_2\right)\hat{\xi}_{oooo},\label{8-5}
\end{eqnarray}
%where $\hat{\xi}_{oooo}$ is an effective energy of interaction, which is equal to the interaction potential $V_0$ for the elementary boson case.
The above results are exact for elementary bosons and only approximate for composite bosons, being valid to lowest order in density only:  indeed, in the case of elementary bosons, the many-body physics is induced by $2\times 2$ interactions while for composite bosons, a quite subtle new set of many body effects arise from fermion exchanges which can exist between more than 2 cobosons.

The single-state case is recovered for $a'$ or $a'' = 0$ in (\ref{8-4}) and for $N_1$ or $N_2 = 0$ in (\ref{8-5}), as expected.  We also see that, since $\hat{\xi}_{oooo}$ must be positive (otherwise the system would suffer a density collapse, its energy decreasing with increasing density), the minimum energy is obtained for macroscopic occupation of the single state $|\phi_0\rangle $.  Consequently, a condensate in just one state is stable---a condensate fragmented into two eigenstates or in a superposition of different eigenstates has a macroscopically higher energy, even when the other states $(o_1,~o_2)$ or $(o',~o'')$ are infinitesimally close to the ground state $o$.

We end with one last comment.  We have here considered the possibility of the condensate to differ from a macroscopically occupied single state through the study of  the Hamiltonian mean value for the states $|\phi_0\rangle ,~|\phi\rangle $ and $|\phi_{12}\rangle $.  Of course, $B^{\dagger N}_o|0\rangle $ is not the true ground state of the system. It however corresponds to this ground state to lowest order in the interaction, as can be seen from the fact that the Hamiltonian mean value of this state has a zero-order term in scattering which is the expected one, namely $NE_o$ (see equation (\ref{8-3})).  It is important to stress, however, that while ``lowest order in the interaction'' is a well-defined concept in the case of elementary particles, for which the Hamiltonian reads $\bar{H}_0 + \bar{V}$, such a concept has no clean meaning for composite bosons, since it is not possible to describe the interaction between cobosons as a potential, and thus define a zero-order Hamiltonian or a zero-order eigenstate.  

{\bf Acknowledgments}. This work has been partially supported by the U.S. National Science Foundation through Grant DMR-0706331. We thank B. Bertoni for contributions to this manuscript.

\section{Appendix: Nozi\`eres's original argument}

For completeness, we here reproduce the original Nozi\`eres's argument on Bose-Einstein condensate \cite{r12}, that there is a macroscopic number of elementary bosons in a single quantum state, not from a difference in the free particle kinetic energy but the exchange part of the interaction energy.  Nozi\`eres calculates this interaction energy, within the Born approximation, namely
\begin{equation}
{\langle \bar{V} \rangle} = \frac{\langle \bar{\phi} | \bar{V} | \bar{\phi}\rangle}{\langle \bar{\phi}  | \bar{\phi} \rangle},
\label{born}
\end{equation}
for a structureless interaction Hamiltonian 
$$
\bar{V} = \frac{V_0}{2}\sum_{\vec{k}_1, \vec{k}_2,\vec{k}_3}\bar{B}^{\dagger}_{\vec{k}_1} \bar{B}^{\dagger}_{\vec{k}_2}  \bar{B}^{ }_{\vec{k}_3}  \bar{B}^{ }_{\vec{k}_1+\vec{k}_2-\vec{k}_3} .
$$
He first considers an elementary boson condensate made of a single sate
$
\  |\bar{\phi}_0\rangle = \bar{B}_0^{{\dagger N}}|0\rangle,
$
the elementary boson operator $\bar{B}_{\vec{k}}$ being such that $[\bar{B}_{\vec{k}},\bar{B}_{\vec{k}'}] = \delta_{\vec{k},\vec{k}'}$.
The interaction energy is then found to be, in the large $N$ limit,
\begin{equation}
{\langle \bar{V} \rangle}_0\simeq \frac{1}{2}V_0N^2.
\label{break1}
\end{equation}
Nozi\`eres then considers a condensate made, not of a single quantum state, but of {\em two} degenerate or nearly-degenerate states, 
$
|\bar{\phi}_{12}\rangle = \bar{B}_{1}^{\dagger {N_1}} \bar{B}_{2}^{\dagger {N_2}}|0\rangle
$,
with the same total number of particles $N_1 + N_2 = N$.
In this fragmented state, the interaction energy is found to be
\begin{equation}
\langle \bar{V} \rangle_{12} =  \langle \bar{V} \rangle_0 + V_0N_1N_2.
\label{condexch}
\end{equation}
For $V_0$ positive, as necessary to prevent a density collapse, this readily shows that it we must pay a {\em macroscopic} amount of energy to break up the condensate into two parts, due to exchange between these two parts, as seen from the $N_1N_2$ prefactor of this additional energy.

Instead of the structureless constant scattering $V_0$ used by Nozi\`eres, we have, in this paper, decided to use a scattering $\bar{\xi}
(^n_m \left. \right.^j_i  )$, which a priori depends on the ``in" and ``out" states. The idea is to make an easier comparison with the results obtained for composite bosons which read in terms of the specific combination of direct and exchange processes given in (\ref{I14}), namely $\hat{\xi}{ (^n_m \left. \right.^j_i  )} = \xi{ (^n_m \left. \right.^j_i  )} - \xi^{in}{ (^n_m \left. \right.^j_i  )}$. This, in particular, allows us to identify the proper effective scattering we must use for bosonized particles. Let us however stress that this identification can only be done for energy conserving processes as in the case of {\em diagonal} processes, i.e., processes in which the ``in" and ``out" states are identical. Indeed, such a scattering  $\hat{\xi}{ (^n_m \left. \right.^j_i  )}$  cannot be used in general because, as $(\xi^{in}{ (^n_m \left. \right.^j_i  )})^*=\xi^{out}{ (^i_j \left. \right.^m_n  )}$, the resulting Hamiltonian would not be Hermitian, due to (\ref{pscatt}) which tells that the ``in" and ``out" scatterings are equal only when energy is conserved. This difficulty is basically linked to the fact that there is no way to have an effective Hamiltonian for bosonized particles which is Hermitian  {\em and} valid for all many-body effects. By taking a real constant $V_0$, Nozi\`eres hides this difficulty.

\end{document}